\documentclass[apj]{emulateapj}
\usepackage{multirow}
\usepackage{longtable}
\usepackage{rotating}

\def\gtorder{\mathrel{\raise.3ex\hbox{$>$}\mkern-14mu
             \lower0.6ex\hbox{$\sim$}}}
\def\ltorder{\mathrel{\raise.3ex\hbox{$<$}\mkern-14mu
             \lower0.6ex\hbox{$\sim$}}}

\slugcomment{Draft of \today}

\shorttitle{Spin Axes and Shape Models of Asteroid Pairs}

\shortauthors{Polishook}

\begin{document}

\title{SPIN AXES AND SHAPE MODELS OF ASTEROID PAIRS: FINGERPRINTS OF YORP AND A PATH TO THE DENSITY OF RUBBLE PILES}

\author{D.~Polishook\altaffilmark{1}
}

\altaffiltext{1}{Department of Earth, Atmospheric, and Planetary Sciences, Massachusetts Institute of Technology, Cambridge, MA 02139, USA}

\begin{abstract}
	An asteroid pair consists of two unbound objects with almost identical heliocentric orbital elements that were formed when a single ``rubble pile" asteroid failed to remain bound against an increasing rotation rate. Models suggest that the pairs' progenitors gained the fast rotation due to the YORP effect. Since it was shown that the spin axis vector can be aligned by the YORP effect, such a behavior should be seen on asteroid pairs, if they were indeed formed by the described mechanism. Alternatively, if the pairs were formed by a collision, the spin axes should have a random direction and small or young bodies might have a tumbling rotation.

	Here I apply the lightcurve inversion method on self-obtained photometric data, in order to derive the rotation axis vectors and shape models of the asteroid pairs 2110, 3749, 5026, 6070, 7343 and 44612. Three asteroids resulted with polar-directed spin axes and three objects with ambiguous results. In addition, the secondary member 44612 presents the same sense of rotation as its primary member 2110, and its spin is not tumbling. Finally, I use a rotational fission model, based on the assumption of an angular momentum conservation, and match it to the measured spin, shape, and mass ratio parameters in order to constrain the density of the primary members in the pairs. Using this method, low density values that are expected from a ``rubble pile" are derived. All these results lead to the conclusion that the disruption of these asteroid pairs was most likely the outcome of the YORP effect that spun-up ``rubble pile" asteroids.


\end{abstract}

\keywords{
Asteroids; Asteroids, rotation; Rotational dynamics; Photometry}

\section{Introduction}
\label{sec:Introduction}

\subsection{Spin evolution mechanisms}
\label{sec:spinevolution}

	The Yarkovsky-ÐO'Keefe-ÐRadzievskii-ÐPaddack (YORP) effect is a thermal torque imposed on asteroids due to the reflection and re-emission of sunlight from the body'Õs asymmetric surface (Rubincam 2000; Bottke et al. 2006). Since the YORP effect is applied by the momentum carried by sunlight photons, it is mainly a function of the asteroid radius $R$ and its heliocentric distance, characterized by its semi major-axis $a$. The resulted change in the spin rate of the asteroid $d \omega/dt$ can be defined by (Scheeres 2007):
\begin{equation}
\frac{d \omega}{dt} = \frac{Y}{2 \pi \rho R^2} \frac{F}{a^2 \sqrt{1-e^2}}
\label{eq:yorp}
\end{equation}
where $\rho$ is the density of the body and $e$ is the eccentricity of its orbit. $F$ is the solar irradiance ($1.361~kW/m^2$, at $1 AU$) modified by the speed of light to derive solar radiation pressure and normalized to a unit distance ($\sim1 \cdot 10^{14} ~kg ~km ~s^{-2}$; Scheeres 2007). $Y$ is a non-dimensional YORP coefficient determined by the asymmetric shape of the asteroid and the obliquity of its rotation. The YORP timescale $\tau _{yorp}$ is defined by:
\begin{equation}
\tau _{yorp} = \frac{\omega}{\mid \frac{d \omega}{dt} \mid}
\label{eq:tauyorp}
\end{equation}

Where $\omega$ is the spin rate. The change $dP/dt$ in the rotation period $P$ can be defined by (Rozitis \& Green 2013):
\begin{equation}
\frac{dP}{dt} = \pm \frac{P^2}{2\pi} \frac{d \omega}{dt} 
\label{eq:dPdt}
\end{equation}

	 Observational studies have shown that the YORP effect can double the spin rate of an asteroid in a relatively short timescale of about a million years for a km-sized near-Earth asteroids (e.g., Lowry et al. 2007, Taylor et al. 2007, Kaasalainen et al. 2007, {\v D}urech et al. 2008, 2012). Such a short timescale makes the YORP effect a very efficient mechanism to control the spins of small bodies (up to diameter of {$\sim10$} km) among the near-Earth asteroids (NEAs) and main-belt asteroids (MBAs) alike, according to their spin distributions (e.g., Pravec et al. 2008, Polishook \& Brosch 2009, Statler et al. 2013). While the rotation of an asteroid can also be spun-up by sub-catastrophic impacts (Paolicchi et al. 2002), the YORP effect seems to be a more robust and efficient process for small-sized asteroids (Marzari et al. 2011, Jacobson et al. 2014).

	The YORP effect is also known to modify the obliquity of asteroids since the re-emitted light has a non-orthogonal component imposed on the spin axis. Hanu{\v s} et al. (2011) found that the latitude distribution of small asteroids (D $< 30$ km) is clustered towards ecliptic poles and explained it by the YORP effect. Slivan (2002) found that asteroids of the Koronis family tend to cluster in two specific states (one in prograde and the second in a retrograde rotation) even though the Koronis family asteroids were formed in a catastrophic collision, and their obliquities and rotation periods should have a random distribution (Asphaug \& Scheeres 1999, Paolicchi et al. 2002). Vokrouhlick{\'y} et al. (2003) showed how the measured obliquities of the Koronis family can be the result of the YORP effect and a spin-orbit resonance. Therefore, asteroids that were formed by a catastrophic collision, within a recent time that is shorter than the timescale of the YORP effect, should present a randomized obliquity distribution rather than a distribution biased for a specific rotation state. The same is true for large asteroids that were formed by catastrophic collisions: since their YORP timescale exceeds the age of the solar system, they present a randomized obliquity distribution (Hanu{\v s} et al. 2013).
	
	Not only do catastrophic collisions randomize the obliquity distribution of the fragments, they also give the resulted asteroids a wide range of rotation periods characterized by complex rotations as was shown by numerical calculations (Asphaug \& Scheeres 1999) and laboratory experiments (Giblin et al. 1998). Complex rotations around a non-principal axis are referred to as {\it tumbling asteroids} (Pravec et al. 2005) and the time $\tau _{damp}$ needed to damp the excited spin into an uniform rotation was derived by Burns \& Safranov (1973) as:
\begin{equation}
\tau_{damp} \sim \frac{\mu Q}{\rho K_3^2R^2 \omega^3}.
\label{eq:damping}
\end{equation}
where $\mu$ is the rigidity of the material the asteroid is composed of, $Q$ is the ratio of the energy contained in the oscillation to the energy lost per cycle (``quality factor"), $\rho$ is the asteroid's density, $R$ is its mean radius and $\omega$ is its spin rate. $K_3^2$ is a dimensionless factor relating to the asteroid's shape. Using reasonable values described by Harris (1994), one can calculate that the damping time for a tumbling km-sized asteroid is $\sim 10^6$ to $\sim 10^8$ years depending on the rotation period. Therefore, asteroids that were formed by a catastrophic collision, within a recent time that is shorter than their damping timescale, should present tumbling rotations rather than a relaxed rotation around a single axis.

\subsection{Asteroid pairs}
\label{sec:pairs}

	{\it Pairs} of asteroids that share almost identical heliocentric orbits were shown by dynamical calculations to originate from a single progenitor and not to be a mere coincidence (Vokrouhlick{\'y} \& Nesvorn{\'y} 2008, 2009, Pravec \& Vokrouhlick{\'y} 2009). Members belonging to the same pair present similar spectral behavior or broadband colors, without a single case of a significant mismatch observed (up until now, similarities were observed among 20 {\it asteroid pairs}: Moskovitz 2012, Duddy et al. 2012, 2013, Polishook et al. 2014, Wolters et al. 2014). This result supports the notion of a single origin for each pair to a significance of almost $5\sigma$.
	
	The formation of {\it asteroid pairs} was explained as the subsequent outcome of the YORP effect: following the spin-up of an asteroid by the YORP effect, an asteroid gains sufficient angular momentum to cross the breakup limit for a strengthless, ``rubble pile'' object (Margot et al. 2002, Scheeres 2007, Walsh et al. 2008, Jacobson \& Scheeres 2011), and the asteroid split into a {\it pair} of asteroids\footnote{In this context, the term ``{\it asteroid pair}" originates from the discovery circumstances of this class of objects, but it also describes an asteroid that broke apart into multiple bodies. This was recently observed by Jewitt et al. (2014) that follow the disintegration of the main belt asteroid {\it P/2013 R3} into, at least, 10 fragments. The main belt asteroids {\it P/2010 A2} and {\it P/2013 P5} probably suffered from the same fate (Jewitt et al. 2010, 2013). } (Pravec et al. 2010). Alternatively, a catastrophic collision could form the pairs (Durda et al. 2004), even though the slow drifting velocity between the components of each pair is less likely in a scenario of a collision (Vokrouhlick{\'y} \& Nesvorn{\'y} 2008).

\subsection{Study goals}
\label{sec:goals}

	Measuring the spin axis of asteroid pairs could disentangle between the two formation models: if the latitude distribution presents high preference for ecliptic poles then the YORP effect is the relevant mechanism for the pairs formation; if the latitude distribution is randomized than collisions shuttered the pairs' progenitors. In addition, a collision event can explain two members of a single asteroid pair that have significantly different spin axes, while the rotational fission model cannot. The model suggested by Pravec et al. (2010) assumed that the spin axes of a pair's members are parallel, and their model matches to the measurements of the pairs' spin periods and mass ratios. This match suggests that their assumption is indeed valid. Here, I am directly measuring the spin axes of asteroid pairs in order to confirm this assumption.

	An important factor is the short time that passed since the pairs formation ($10^4$ to $10^6$ years; Vokrouhlick{\'y} \& Nesvorn{\'y} 2008, Pravec et al. 2010, Polishook et al. 2014), that is referred here as the ``dynamical age". If this time is shorter than the YORP timescale than we can assume that the YORP effect did not alter the spin vector since the fission event occur.

	Same goes for the internal damping timescale. If an asteroid pair has a damping timescale that is longer than its dynamical age, and if the spin does not presents a tumbling nature, then the pair was not formed by a catastrophic collision. This is especially true for the smaller members of the pairs that have longer damping times.
	
	In addition, knowing the sense of rotation of {\it asteroid pairs} is important for calculating the time passed since their formation: the sense of rotation determines the sign of the Yarkovsky drag (Bottke et al. 2006) imposed on the asteroid and it modifies its heliocentric orbit. This is an essential parameter when integrating backward an asteroid's orbital elements to derive the dynamical age of the pair.
	
	And finally, deriving the spin axis by the lightcurve inversion method also constrains the shape model of the asteroid. Assuming a pair formed by the YORP effect, the shape and spin parameters can further constrain the density of the rotational-fissioned asteroid using an angular momentum conservation model as describe below. If the derived density values are low compared to the density of the material the asteroids consist of, than we can conclude these asteroids have a strengthless, {\it rubble-pile} structure, an essential characteristic for the rotational-fission mechanism to take place.

\section{Observations, reduction, measurements and calibration}
\label{sec:observations}

\subsection{Observations from the Wise Observatory}
\label{sec:wise}

	I collected photometric data on six asteroids during more than 110 nights between 2007 to 2014 (fortunately, most observations were performed in a remote operation mode). All observations were performed using the two telescopes of the Wise Observatory in Israel (code: 097): the 1-m Ritchey-ChrŽtien telescope; and the 0.46-m Centurion telescope (Brosch et al. 2008). The 1-m telescope is equipped with a cryogenically-cooled Princeton Instruments (PI) CCD. At the f/7 focus of the telescope this CCD covers a field of view of 13'x13' with 1340x1300 pixels (0.58'' per pixel, unbinned). The 0.46-m is equipped with a wide-field SBIG STL-6303E CCD (75'x55' with 3072x2048 pixels, 1.47'' per pixel, unbinned). Most observations were performed without filters ("Clear"), but some were done using Cousins {\it R}. To achieve a point-like FWHM for the moving targets (at a seeing value of $\sim$2.5 arcsec), exposure times of $60-240$ seconds were used, all with an auto-guider. The observational circumstances are summarized in Table~\ref{tab:ObsCircum1}.

The images were reduced in a standard way using bias and dark subtraction, and were divided by a normalized flatfield image. Astrometric solutions were obtained using the {\it PinPoint} software\footnote{www.dc3.com} and the asteroids were identified against the MPC web database\footnote{http://www.minorplanetcenter.org/iau/mpc.html}. {\it IRAF}'s {\it phot} function was used for the photometry. Apertures with four-pixel radii were chosen to minimize photometric errors. The mean sky value was measured using an annulus of 10 pixels wide and inner radius of 10 pixels around the asteroid.

The photometric values were calibrated to a differential magnitude level using local comparison stars measured on every image using the same method as the asteroid. Variable stars were removed at a second iteration. The brightness of the comparison stars remained constant to $\pm$0.02 mag. A photometric shift was calculated for each image compared to a good reference image, using the local comparison stars. The photometric data of asteroid 44612 were also calibrated against the standard magnitude scale using Landolt equatorial standards (Landolt 1992). The asteroid data were corrected for light-travel time and the magnitudes were reduced to one AU distance from the Sun and the Earth (Bowell et al. 1989). Clear outliers were excluded manually. Refer to Polishook and Brosch (2008, 2009) for detailed description of the photometric procedures of observation, reduction and calibration using the 1-m and the 0.46-m, respectively. Some of the observations were previously published by Pravec et al. (2010), Polishook et al. (2011) and Vokrouhlick{\'y} et al. (2011).

\begin{deluxetable*}{cccccccccccc}
\tablecolumns{12}
\tablewidth{0pt}
\tablecaption{Observation circumstances}
\tablehead{
\colhead{Asteroid} &
\colhead{Date} &
\colhead{Telescope} &
\colhead{Filter} &
\colhead{Time span} &
\colhead{N} &
\colhead{${\it r}$} &
\colhead{${\it \Delta}$} &
\colhead{${\it \alpha}$} &
\colhead{${\it \lambda_{PAB}}$} &
\colhead{${\it \beta_{PAB}}$} &
\colhead{Ref.} \\
\colhead{}          &
\colhead{}          &
\colhead{}          &
\colhead{}          &
\colhead{[hour]}    &
\colhead{}          &
\colhead{[AU]} &
\colhead{[AU]} &
\colhead{[deg]} &
\colhead{[deg]} &
\colhead{[deg]} &
\colhead{}
}
\startdata
2110 & 20081219 & 0.46m & Clear & 2.77 & 33 & 2.27 & 1.29 & 3.73 & 82.2 & -1.2 & Pravec et al. \\
 & 20081227 & 0.46m & Clear & 5.86 & 71 & 2.29 & 1.33 & 7.85 & 82.8 & -1.2 & Pravec et al. \\
 & 20090102 & 0.46m & Clear & 7.01 & 89 & 2.3 & 1.37 & 10.78 & 83.3 & -1.1 & Pravec et al. \\
 & 20090103 & 0.46m & Clear & 7.19 & 62 & 2.3 & 1.38 & 11.23 & 83.4 & -1.1 & Pravec et al. \\
 & 20111027 & 0.46m & Clear & 2.37 & 48 & 1.97 & 0.98 & 2.09 & 36.2 & -1.5 & --- \\
 & 20111028 & 0.46m & Clear & 2.82 & 56 & 1.97 & 0.98 & 1.62 & 36.4 & -1.5 & --- \\
 & 20130202 & 0.46m & Clear & 2.23 & 28 & 2.59 & 1.72 & 12.93 & 162.5 & 0.4 & --- \\
 & 20130203 & 0.46m & Clear & 2.15 & 27 & 2.59 & 1.72 & 12.57 & 162.5 & 0.4 & --- \\
 & 20130207 & 0.46m & Clear & 2.72 & 46 & 2.59 & 1.69 & 11.03 & 162.7 & 0.4 & --- \\
 & 20130208 & 0.46m & Clear & 2.28 & 37 & 2.59 & 1.68 & 10.62 & 162.8 & 0.4 & --- \\
 & 20130211 & 0.46m & Clear & 3.98 & 46 & 2.59 & 1.66 & 9.45 & 162.9 & 0.4 & --- \\
 & 20130212 & 0.46m & Clear & 7.19 & 71 & 2.59 & 1.65 & 8.98 & 162.9 & 0.4 & --- \\

3749 & 20070715 & 1m & R & 1.11 & 12 & 2.37 & 1.87 & 24.27 & 352.6 & 4.4 & Polishook et al. \\
 & 20070719 & 1m & R & 3.62 & 37 & 2.36 & 1.82 & 23.83 & 353.3 & 4.6 & Polishook et al. \\
 & 20070723 & 1m & R & 3.55 & 31 & 2.36 & 1.77 & 23.3 & 354 & 4.7 & Polishook et al. \\
 & 20070811 & 1m & R & 0.9 & 12 & 2.34 & 1.55 & 19.3 & 356.7 & 5.3 & Polishook et al. \\
 & 20070812 & 1m & R & 3.01 & 35 & 2.34 & 1.54 & 19.03 & 356.8 & 5.3 & Polishook et al. \\
 & 20070816 & 1m & R & 3.12 & 23 & 2.34 & 1.51 & 17.87 & 357.3 & 5.4 & Polishook et al. \\
 & 20071004 & 1m & R & 5.19 & 60 & 2.28 & 1.3 & 6.8 & 1.8 & 6.4 & Polishook et al. \\
 & 20071005 & 1m & R & 6.95 & 71 & 2.28 & 1.3 & 7.19 & 1.9 & 6.4 & Polishook et al. \\
 & 20071006 & 1m & R & 7.11 & 44 & 2.27 & 1.3 & 7.68 & 2 & 6.4 & Polishook et al. \\
 & 20071105 & 1m & R & 2.75 & 20 & 2.24 & 1.47 & 19.88 & 5.3 & 6.4 & Polishook et al. \\
 & 20071213 & 1m & R & 4.27 & 14 & 2.19 & 1.85 & 26.53 & 12.3 & 6 & Polishook et al. \\
 & 20090316 & 0.46m & Clear & 1.76 & 30 & 2.15 & 1.17 & 7.8 & 186.5 & -6.5 & --- \\
 & 20090317 & 0.46m & Clear & 0.34 & 7 & 2.15 & 1.17 & 7.35 & 186.6 & -6.5 & --- \\
 & 20090321 & 0.46m & Clear & 2.55 & 53 & 2.15 & 1.17 & 5.72 & 187.1 & -6.6 & --- \\
 & 20090328 & 0.46m & Clear & 0.95 & 20 & 2.16 & 1.17 & 4.38 & 187.8 & -6.7 & --- \\
 & 20100609 & 0.46m & Clear & 1.03 & 11 & 2.48 & 1.87 & 21.8 & 311.4 & 0.7 & --- \\
 & 20100709 & 0.46m & Clear & 2.15 & 26 & 2.47 & 1.57 & 13.91 & 314.6 & 1.5 & --- \\
 & 20100717 & 0.46m & Clear & 1.72 & 24 & 2.46 & 1.51 & 10.69 & 315.2 & 1.8 & --- \\
 & 20100718 & 0.46m & Clear & 2.41 & 36 & 2.46 & 1.51 & 10.27 & 315.3 & 1.8 & --- \\
 & 20100719 & 0.46m & Clear & 5.34 & 47 & 2.46 & 1.5 & 9.86 & 315.3 & 1.8 & --- \\
 & 20100804 & 0.46m & Clear & 4.87 & 72 & 2.45 & 1.44 & 2.56 & 316.3 & 2.3 & --- \\
 & 20100805 & 0.46m & Clear & 6.92 & 48 & 2.45 & 1.44 & 2.18 & 316.3 & 2.3 & --- \\
 & 20100806 & 0.46m & Clear & 3.19 & 47 & 2.45 & 1.44 & 1.8 & 316.4 & 2.3 & --- \\
 & 20100807 & 1m & Clear & 5.33 & 108 & 2.45 & 1.44 & 1.52 & 316.5 & 2.4 & --- \\
 & 20100808 & 1m & Clear & 5.25 & 104 & 2.45 & 1.44 & 1.36 & 316.5 & 2.4 & --- \\
 & 20100809 & 0.46m & Clear & 1.48 & 26 & 2.45 & 1.44 & 1.37 & 316.6 & 2.4 & --- \\
 & 20100810 & 0.46m & Clear & 0.49 & 8 & 2.45 & 1.44 & 1.53 & 316.6 & 2.4 & --- \\
 & 20100812 & 0.46m & Clear & 1.26 & 20 & 2.45 & 1.44 & 2.31 & 316.7 & 2.5 & --- \\
 & 20100813 & 1m & Clear & 6.91 & 76 & 2.45 & 1.44 & 2.66 & 316.8 & 2.5 & --- \\
 & 20100814 & 1m & Clear & 4.61 & 83 & 2.45 & 1.44 & 3.1 & 316.9 & 2.6 & --- \\
 & 20100817 & 0.46m & Clear & 2.48 & 32 & 2.45 & 1.44 & 4.51 & 317 & 2.6 & --- \\
 & 20111219 & 0.46m & Clear & 3.35 & 79 & 1.99 & 1.14 & 18.74 & 116.2 & 0.8 & --- \\
 & 20111228 & 0.46m & Clear & 2.09 & 49 & 1.99 & 1.08 & 14.47 & 117.8 & 0.4 & --- \\
 & 20120102 & 0.46m & Clear & 2.42 & 44 & 1.99 & 1.05 & 11.79 & 118.7 & 0.2 & --- \\
 & 20120121 & 0.46m & Clear & 3.33 & 84 & 1.99 & 1.01 & 0.6 & 121.7 & -0.7 & --- \\
 & 20120122 & 0.46m & Clear & 3.67 & 37 & 1.99 & 1.01 & 0.6 & 121.8 & -0.8 & --- \\
 & 20120123 & 0.46m & Clear & 7.21 & 193 & 1.99 & 1.01 & 1.09 & 121.9 & -0.8 & --- \\
 & 20120220 & 0.46m & Clear & 7.15 & 162 & 2 & 1.12 & 16.79 & 126.5 & -2 & --- \\

5026 & 20090715 & 0.46m & Clear & 1.42 & 30 & 1.8 & 0.96 & 24.69 & 325 & 1.2 & Pravec et al. \\
 & 20090716 & 0.46m & Clear & 1.66 & 33 & 1.8 & 0.95 & 24.29 & 325.4 & 1.2 & Pravec et al. \\
 & 20090717 & 0.46m & Clear & 4.06 & 81 & 1.8 & 0.94 & 23.95 & 325.6 & 1.3 & Pravec et al. \\
 & 20090813 & 0.46m & Clear & 3 & 41 & 1.8 & 0.82 & 10.81 & 333 & 2.5 & Pravec et al. \\
 & 20090814 & 0.46m & Clear & 1.14 & 24 & 1.81 & 0.82 & 10.26 & 333.2 & 2.5 & Pravec et al. \\
 & 20110306 & 1m & Clear & 3.3 & 18 & 2.93 & 2.23 & 15.87 & 121.1 & -0.3 & --- \\
 & 20110308 & 1m & Clear & 5.51 & 62 & 2.93 & 2.25 & 16.28 & 121.2 & -0.3 & --- \\
 & 20120319 & 0.46m & Clear & 3.6 & 50 & 2.6 & 1.68 & 10.25 & 200.5 & -5 & --- \\
 & 20120323 & 0.46m & Clear & 0.87 & 16 & 2.59 & 1.65 & 8.72 & 200.7 & -5 & --- \\
 & 20120416 & 0.46m & Clear & 2.98 & 45 & 2.54 & 1.54 & 3.91 & 201.8 & -5.2 & --- \\
 & 20120419 & 0.46m & Clear & 6.13 & 95 & 2.53 & 1.54 & 5.04 & 201.9 & -5.2 & --- \\
 & 20120512 & 0.46m & Clear & 0.32 & 6 & 2.47 & 1.6 & 14.73 & 203.2 & -5.2 & --- \\
 & 20120620 & 0.46m & Clear & 1.44 & 19 & 2.37 & 1.9 & 24.55 & 207.8 & -4.9 & --- \\
 & 20120621 & 0.46m & Clear & 1.47 & 10 & 2.37 & 1.91 & 24.68 & 208 & -4.8 & --- \\
 & 20131027 & 0.46m & Clear & 5.89 & 73 & 2.19 & 1.24 & 10.21 & 49.9 & 5.4 & --- \\
 & 20131030 & 0.46m & Clear & 2.9 & 45 & 2.2 & 1.23 & 8.57 & 50.3 & 5.4 & --- \\
 & 20131112 & 0.46m & Clear & 2.23 & 35 & 2.23 & 1.25 & 3.35 & 51.5 & 5.3 & --- \\
 & 20140106 & 1m & Clear & 5.79 & 76 & 2.38 & 1.74 & 21.14 & 57.6 & 4.3 & --- \\
 \enddata
\tablecomments{Columns: asteroids' designations, date of observation, telescope, filter, hourly time span, number of images, heliocentric and geocentric distances, phase angle, longitude and latitude of the Phase Angle Bisector (PAB), reference for previously published data: Pravec et al. 2010, Polishook et al. 2011. All observations were performed by the author at the Wise Observatory}
\label{tab:ObsCircum1}
\end{deluxetable*}

\begin{deluxetable*}{cccccccccccc}
\setcounter{table}{1}
\tablecolumns{12}
\tablewidth{0pt}
\tablecaption{Observation circumstances - continue}
\tablehead{
\colhead{Asteroid} &
\colhead{Date} &
\colhead{Telescope} &
\colhead{Filter} &
\colhead{Time span} &
\colhead{N} &
\colhead{${\it r}$} &
\colhead{${\it \Delta}$} &
\colhead{${\it \alpha}$} &
\colhead{${\it \lambda_{PAB}}$} &
\colhead{${\it \beta_{PAB}}$} &
\colhead{Ref.} \\
\colhead{}          &
\colhead{}          &
\colhead{}          &
\colhead{}          &
\colhead{[hour]}    &
\colhead{}          &
\colhead{[AU]} &
\colhead{[AU]} &
\colhead{[deg]} &
\colhead{[deg]} &
\colhead{[deg]} &
\colhead{}
}
\startdata
6070 & 20090721 & 0.46m & Clear & 2.48 & 31 & 1.98 & 1.26 & 26.28 & 342.7 & -3.7 & Pravec et al. \\
 & 20090723 & 0.46m & Clear & 1.78 & 27 & 1.98 & 1.24 & 25.86 & 343.2 & -3.7 & Pravec et al. \\
 & 20090724 & 0.46m & Clear & 3.81 & 57 & 1.98 & 1.23 & 25.65 & 343.5 & -3.7 & Pravec et al. \\
 & 20090725 & 0.46m & Clear & 1.96 & 23 & 1.97 & 1.22 & 25.41 & 343.7 & -3.7 & Pravec et al. \\
 & 20090817 & 0.46m & Clear & 1.88 & 28 & 1.94 & 1.03 & 17.94 & 349.1 & -4.1 & Pravec et al. \\
 & 20090920 & 0.46m & Clear & 6.58 & 130 & 1.9 & 0.9 & 3.73 & 356.3 & -4.2 & Pravec et al. \\
 & 20091020 & 0.46m & Clear & 6.36 & 67 & 1.89 & 0.99 & 18.55 & 2.8 & -4 & Pravec et al. \\
 & 20091023 & 0.46m & Clear & 4.78 & 71 & 1.89 & 1.01 & 19.85 & 3.5 & -4 & Pravec et al. \\
 & 20091120 & 0.46m & Clear & 3.82 & 56 & 1.88 & 1.24 & 28.29 & 11 & -3.4 & Pravec et al. \\
 & 20091122 & 0.46m & Clear & 1.35 & 20 & 1.89 & 1.26 & 28.63 & 11.6 & -3.4 & Pravec et al. \\
 & 20091221 & 0.46m & Clear & 1.26 & 22 & 1.9 & 1.56 & 31.1 & 21 & -2.7 & --- \\
 & 20110303 & 0.46m & Clear & 7.92 & 95 & 2.75 & 1.77 & 1.92 & 161 & 3.7 & Vokrouhlick{\'y} et al. \\
 & 20110307 & 1m & Clear & 8.25 & 101 & 2.76 & 1.77 & 3.12 & 161 & 3.7 & Vokrouhlick{\'y} et al. \\
 & 20110401 & 1m & Clear & 1.47 & 21 & 2.79 & 1.93 & 12.61 & 161.8 & 3.7 & Vokrouhlick{\'y} et al. \\
 & 20110401 & 0.46m & Clear & 4.54 & 45 & 2.79 & 1.93 & 12.56 & 161.8 & 3.7 & Vokrouhlick{\'y} et al. \\
 & 20120525 & 0.46m & Clear & 5.42 & 55 & 2.65 & 1.64 & 0.65 & 245.3 & 1.2 & --- \\
 & 20120711 & 0.46m & Clear & 0.78 & 13 & 2.56 & 1.81 & 18.38 & 247.9 & 0.6 & --- \\
 & 20120712 & 0.46m & Clear & 1.04 & 18 & 2.56 & 1.82 & 18.64 & 248 & 0.5 & --- \\
 & 20120714 & 0.46m & Clear & 2.06 & 25 & 2.56 & 1.84 & 19.15 & 248.2 & 0.5 & --- \\
 & 20120716 & 0.46m & Clear & 1.7 & 26 & 2.55 & 1.86 & 19.63 & 248.4 & 0.5 & --- \\
 & 20131025 & 0.46m & Clear & 5.88 & 87 & 1.97 & 1.15 & 21.76 & 65.9 & -1.8 & --- \\
 & 20131028 & 0.46m & Clear & 2.03 & 33 & 1.97 & 1.13 & 20.47 & 66.6 & -1.7 & --- \\
 & 20140103 & 0.46m & Clear & 0.87 & 14 & 2.1 & 1.22 & 15.9 & 77.2 & -0.1 & --- \\
 & 20140106 & 0.46m & Clear & 7.86 & 92 & 2.11 & 1.25 & 17.18 & 77.7 & 0 & --- \\

7343 & 20090718 & 0.46m & Clear & 3.37 & 45 & 2.26 & 1.25 & 3.63 & 299.6 & -4.7 & Pravec et al. \\
 & 20090723 & 0.46m & Clear & 4.28 & 38 & 2.25 & 1.24 & 2.99 & 300.1 & -4.6 & Pravec et al. \\
 & 20090724 & 0.46m & Clear & 0.63 & 9 & 2.25 & 1.24 & 3.14 & 300.2 & -4.6 & Pravec et al. \\
 & 20090725 & 0.46m & Clear & 2.67 & 31 & 2.25 & 1.23 & 3.38 & 300.3 & -4.6 & Pravec et al. \\
 & 20090728 & 0.46m & Clear & 1.31 & 20 & 2.24 & 1.23 & 4.44 & 300.6 & -4.5 & Pravec et al. \\
 & 20110307 & 0.46m & Clear & 5.11 & 80 & 2.22 & 1.29 & 11.7 & 147.4 & 2.7 & --- \\
 & 20120620 & 0.46m & Clear & 2.99 & 28 & 2.41 & 1.4 & 3.23 & 267.1 & -5 & --- \\
 & 20120621 & 0.46m & Clear & 3.32 & 35 & 2.41 & 1.4 & 3.51 & 267.2 & -5 & --- \\
 & 20131104 & 0.46m & Clear & 3.56 & 52 & 1.92 & 1.26 & 27.57 & 89.2 & 4.7 & --- \\
 & 20140108 & 0.46m & Clear & 9.63 & 155 & 1.99 & 1.02 & 6.33 & 101.3 & 5.2 & --- \\
 & 20140127 & 0.46m & Clear & 0.59 & 9 & 2.01 & 1.12 & 15.77 & 104.3 & 5 & --- \\

44612 & 20091022 & 1m & Clear & 4.73 & 54 & 2.01 & 1.06 & 11.25 & 45.4 & -1.5 & Pravec et al. \\
 & 20091024 & 1m & Clear & 2.07 & 29 & 2.01 & 1.05 & 10.08 & 45.7 & -1.5 & Pravec et al. \\
 & 20120811 & 0.46m & Clear & 5.27 & 82 & 1.81 & 0.86 & 16.96 & 339.3 & -0.3 & --- \\
 & 20120816 & 0.46m & Clear & 5.59 & 86 & 1.81 & 0.84 & 14.21 & 340.6 & -0.4 & --- \\
 & 20120817 & 0.46m & Clear & 4.05 & 52 & 1.81 & 0.84 & 13.58 & 340.8 & -0.4 & --- \\
 & 20120820 & 0.46m & Clear & 5.82 & 82 & 1.81 & 0.83 & 11.84 & 341.5 & -0.4 & --- \\
 & 20120822 & 0.46m & Clear & 5.46 & 100 & 1.81 & 0.82 & 10.65 & 342 & -0.5 & --- \\
 & 20120825 & 0.46m & Clear & 7.12 & 84 & 1.81 & 0.82 & 8.79 & 342.7 & -0.5 & --- \\
 & 20120827 & 0.46m & Clear & 6.08 & 77 & 1.81 & 0.81 & 7.48 & 343.2 & -0.5 & --- \\
 & 20120904 & 0.46m & Clear & 5.9 & 101 & 1.81 & 0.81 & 2.29 & 345.1 & -0.6 & --- \\
 & 20120906 & 0.46m & Clear & 7.15 & 80 & 1.82 & 0.81 & 1.08 & 345.5 & -0.6 & --- \\
 & 20120907 & 0.46m & Clear & 2.3 & 42 & 1.82 & 0.81 & 0.65 & 345.7 & -0.6 & --- \\
 & 20120909 & 0.46m & Clear & 1.73 & 27 & 1.82 & 0.81 & 1.16 & 346.2 & -0.7 & --- \\
 & 20120911 & 0.46m & Clear & 2.23 & 39 & 1.82 & 0.81 & 2.55 & 346.7 & -0.7 & --- \\
 & 20120914 & 0.46m & Clear & 5.02 & 85 & 1.82 & 0.82 & 4.38 & 347.3 & -0.7 & --- \\
 & 20120917 & 0.46m & Clear & 7.12 & 98 & 1.82 & 0.83 & 6.32 & 348 & -0.7 & --- \\
 & 20120920 & 0.46m & Clear & 1.37 & 22 & 1.82 & 0.84 & 8.15 & 348.7 & -0.8 & --- \\
 & 20120923 & 0.46m & Clear & 4.55 & 73 & 1.83 & 0.85 & 10.05 & 349.4 & -0.8 & --- \\
 & 20121009 & 0.46m & Clear & 5.76 & 70 & 1.84 & 0.94 & 18.54 & 353.2 & -0.9 & --- \\
 & 20121010 & 0.46m & Clear & 1.73 & 20 & 1.84 & 0.94 & 19.01 & 353.5 & -0.9 & --- \\
 & 20140207 & 0.46m & Clear & 5.95 & 54 & 2.58 & 1.6 & 1.73 & 142.4 & 0 & ---
\enddata
\tablecomments{Columns: asteroids' designations, date of observation, telescope, filter, hourly time span, number of images, heliocentric and geocentric distances, phase angle, longitude and latitude of the Phase Angle Bisector (PAB), reference for previously published data: Pravec et al. 2010, Vokrouhlick{\'y} et al. 2011. All observations were performed by the author at the Wise Observatory}
\label{tab:ObsCircum2}
\end{deluxetable*}

\subsection{Sparse lightcurves from large-scale surveys}
\label{sec:surveys}

In order to expand the photometric coverage of the asteroids I also used data taken by large-scale surveys. These surveys usually obtain only a few measurements of an asteroid per night, basically for astrometric needs, therefore they result in ``sparse" lightcurves (as opposed to ``dense" lightcurves). However, due to the systematic observation algorithm of these surveys, the asteroids are observed on a wide range of apparitions and phase angles. For example, using data from the Catalina Sky Survey, the observation period of (5026) Martes was stretched backward from 2009 to 1999. Because the photometric calibration of these surveys is at the order of 0.1 mag I followed Hanu{\v s} et al. (2011) and used data from the two surveys that have the best photometric calibration record: U.S. Naval Observatory (689) and Catalina Sky Survey (703; see Table I at Hanu{\v s} et al. 2011). The data were downloaded from the MPC website\footnote{$http://www.minorplanetcenter.net/db\_search$} and were corrected for light-travel time, and heliocentric and geocentric distances of 1 AU. The data from the surveys were treated as relatively calibrated data, and each night and field of view was treated separately.

\section{Analysis}
\label{sec:analysis}
In this section I describe the steps involved in the analysis of the photometric data: deriving the synodic period from data of a single apparition; matching the dataset to a sidereal period; mapping the entire space of spin axis coordinates for the best solution based on the matched sidereal period; constructing the shape model that is constrained from the best fitted spin axis and checking for its physical feasibility; applying the rotational-fission model with the derived physical parameter and deriving constraints on the asteroid's density.

\subsection{Synodic rotation period}
\label{sec:synodic}
I find the synodic rotation period of the asteroids by fitting a second-order ($N_k=2$) Fourier series to the reduced measurements using:
\begin{eqnarray}
M(t) & = & \sum_{k=1,2}^{N_k}{B_k\sin[\frac{2\pi k}{P}(t-t_0)]+C_k\cos[\frac{2\pi k}{P}(t-t_0)]} \cr
  & & + ZP ,
\label{eq:folding}
\end{eqnarray}

where $B_k$ and $C_k$ are the Fourier coefficients, $P$ is the rotation period, $M(t)$ is the photometric data at time $t$ and $t_0$ is an arbitrary epoch. $ZP$ is a constant needed to bring a set of observations from a single night into the same magnitude scale of datasets from other nights. Measurements from different apparitions were examined separately. For a given {\it P}, this yields a set of linear equations, that is solved using least-squares minimization to obtain the free parameters. The frequency with the minimal $\chi^2$ is chosen as the most likely period. The error in the best-fitting frequency is determined by the range of periods with $\chi^2$ smaller than the minimum $\chi^2$ + $\Delta\chi^2$, where $\Delta\chi^2$ is calculated from the inverse $\chi^2$ distribution assuming $1 + 2k$ degrees of freedom. The calculation is performed for a set of trial frequencies centered on the rotation frequency as measured roughly by eye, for assuming two lightcurve peaks per rotation.

In a single case (44612), the asteroid was repeatedly sampled over a wide range of phase angles (0 to 20 degrees). Following the calibration of its photometric data to a standard magnitude level, the data were matched to the {\it H-G} system (Bowell et al. 1989) in order to derive the absolute magnitude of the asteroid. Polishook et al. (2008) describes this fitting in great detail.

\subsection{The lightcurve inversion method}
\label{sec:inversion}
	I use the lightcurve inversion method to constrain the sidereal rotation period with the coordinates of the spin axis, the shape model and the scattering law parameters. This method, developed by Kaasalainen and Torppa (2001) and Kaasalainen et al. (2001), and implemented as software by Josef {\v D}urech\footnote{http://astro.troja.mff.cuni.cz/projects/asteroids3D/} ({\v D}urech et al. 2010) and Josef Hanu{\v s} (Hanu{\v s} et al. 2011) scans the space of free parameters and gives the optimized solution with the minimal $\chi^2$. Experience shows that the best results of this method are achieved when an asteroid is observed on multiple viewing geometries, when its surface is covered from as many directions as possible. This can be performed only on multiple apparitions and phase angles, as graphically displayed for each asteroid in section~\ref{sec:results}.

	In order to determine the statistical uncertainty of the solution I follow the method describe by Vokrouhlick{\'y} et al. (2011). They confronted the problems of photometric uncertainties that may not obey Gaussian statistics and that systematic and modeling errors dominate over the random ones. The $\chi^2$ normalized by the degrees of freedom $\nu$ is used to obtain $\chi^2$ values in the order of one. $\chi^2$-isocontours correspond to solutions with a {\it R} percent larger $\chi^2$ than the global minimum are used to set the limit for statistically acceptable solutions. The rate {\it R} is a function of the degrees of freedom, since a $\chi^2$ distribution with $\nu$ degrees of freedom has a variance of $2\nu$ (Press et al. 2007):
	
\begin{equation}
R = \sigma \frac{\sqrt{2\nu}}{\nu}.
\label{eq:deltaChi2}
\end{equation}

	The degree of freedom $\nu$ is the number of measured points minus the number of free parameters, which is about 50 for the model described here. Therefore, the uncertainty reduced with increasing number of measurements. A higher value for $\sigma$ (3 rather than 1) was chosen if the uncertainty was low enough for a significant result. The $\sigma$ values used to define the uncertainty of each asteroid are detailed in section~\ref{sec:results}.

\subsection{Sidereal rotation period}
\label{sec:sidereal}

\begin{figure}
\centerline{\includegraphics[width=8.5cm]{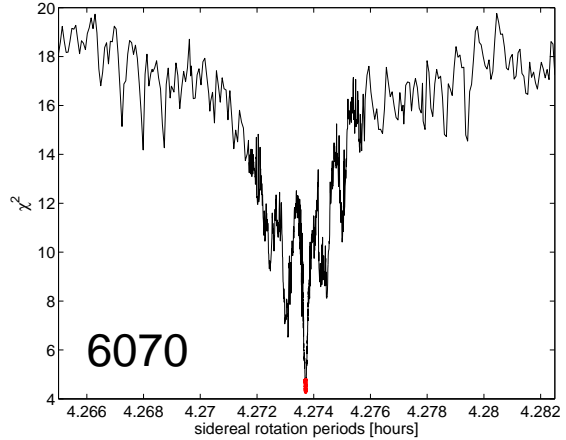}}
\caption{An example for deriving the sidereal period: the $\chi^2$ values for a range of sidereal periods for (6070) {\it Rheinland}. The red line at the bottom marks the uncertainty range that corresponds to $3\sigma$ above the global minimum.
\label{fig:RheinlandSid}}
\end{figure}

At the first step only the sidereal period is being searched for since the other free parameters (e.g. spin axis, shape) are highly sensitive to it, and because mapping the entire range of possible periods is time consuming. Therefore, the code increases the resolution while the range of searched periods is being decreased. In this step, the spin axis, shape and scattering parameters are scanned with low resolution. A preliminary sidereal period is determined as the one with the minimal $\chi^2$ and the uncertainty is defined by all the solutions with {\it R} percent larger $\chi^2$ than the global minimum, as described above (e.g., Fig.~\ref{fig:RheinlandSid}).

In some cases the uncertainty was too high to match the data to a single sidereal period (more than one local minimum was smaller than {\it R} percent larger $\chi^2$ than the global minimum). In such cases all of the sidereal period solutions within the uncertainty were used as input for the spin axis search, as long as the number of solutions was relatively small. The preliminary sidereal period was further tuned in later steps when the spin axis coordinates were scrutinized in higher resolution.

\subsection{Spin axis}
\label{sec:axis}
	Using the derived preliminary sidereal period(s) I apply the lightcurve inversion method on the entire range of the spin axis coordinates using a fixed interval of $5$ degrees. An alternative running mode, where the code is ``free" to choose the best coordinates within the grid (by that resulting with the exact coordinates) is also tested; yet the differences between the two modes are minor for this limited photometric data. The spin axis coordinates are determined by all solutions with $\chi^2$ smaller than {\it R} percent larger $\chi^2$ than the global minimum (Eq.~\ref{eq:deltaChi2}). This uncertainty area is marked by a contour on a longitude-latitude map of solutions. In order to check the validity of the result, the code was run with similar values of sidereal period within its derived error range. If the spin axis solution is similar, we can be confident about the derived solution.
	
	If the spin axis was checked against more than one sidereal period value, resulting with more than one $\chi^2$ maps of spin axes, I use the minimal $\chi^2$ of {\bf all} spin axis solution maps as the best spin axis solution and determine the uncertainty range by it. This allows the rejection of some of the sidereal periods that were obtained in the previous low-resolution step.

\subsection{Shape model}
\label{sec:shape}
	The lightcurve inversion code constructs the shape models that are constrained to the possible spin axis coordinates, using $\sim$2000 triangular facets. For each shape model a triaxial shape (${\it a}\geq{\it b}\geq{\it c}$) was matched by applying a rotation matrix around the shortest physical axis {\it c} while the {\it a} and {\it b} axes are defined as two orthogonal axes on the plain that is orthogonal to the shortest physical axis {\it c}, with {\it a} being the longest axis. Any further interpretation of the shape models (e.g., looking for the area of fission) is premature, with the currently limited photometric data.

	Each shape model is tested for its physical feasibility by two tests: {\it 1)} measuring the angular distance between the axis of maximal moment of inertia to the shortest physical axis of the shape (axis {\it c}). Because the observed asteroids do not have excited rotations and do not ÒtumbleÓ around many axes, the spin axis is expected to be very close to, if not similar, to the shortest physical axis. A tolerance of up to 40 degrees between the shortest physical axis to the axis of maximal moment of inertia was allowed and a ratio between them do not exceed 1.1 (see the rationality in Hanu{\v s} et al. 2011). {\it 2)} the ratio between the triaxial axes of each shape model, {\it a}/{\it b} and {\it b}/{\it c} is calculated. Since the {\it c} axis is parallel to the spin axis and to the axis of maximal moment of inertia, we can assume it to be shorter than the {\it a} and the {\it b} axes. Shape models and their constrained spin axes and sidereal periods that do not fulfill these conditions were rejected.

\subsection{Other free parameters}
\label{sec:other}
The lightcurve inversion code also assumes an asteroid with a convex shape, and a simple empirical scattering law, since it was shown by Kaasalainen's papers that concavities have minimal fingerprints on lightcurves and that the role of the scattering law is very minor compared to the spin axis vector, period and the shape of the photometric convex hull.

\subsection{Testing the analysis method}
\label{sec:test}
	To increase our confidence in the fitted solutions, I run three test studies to check if the solutions are not a mere coincidence: i) the flux values were replaced with randomized values, while keeping the same number of points, and without exceeding the original range of flux values; ii) a spin coordinates solution was search for with a random value for the sidereal period; iii) A single set of dense measurements (data from one night taken at the Wise Observatory) was removed from the entire set, in order to see minor changes in the output. An additional test includes running the code excluding the data from the large-scale surveys.

	In the first two tests, the $\chi^2$ map was flat without any significant solution. These tests confirm that the photometric data can be fitted to a specific solution. The third test always displayed a similar $\chi^2$ map with only larger uncertainty areas. The usage of the sparse lightcurves from the surveys made it possible to reject some sidereal periods and axis solutions and decrease the uncertainty of possible solutions. To conclude, the different tests support the strength of the derived solutions, and we can be certain they are not a mere coincidence.

\subsection{Determine the uncertainty and the statistical significance of the results}
\label{sec:statistics}
	The values I report here for the sidereal rotation period, longitude and latitude of the spin axis, the {\it a}/{\it b} and {\it b}/{\it c} ratios, are the average values from the models that pass the physical feasibility test (section~\ref{sec:shape}) and have a $\chi^2$ value smaller than {\it R} percent larger $\chi^2$ than the global minimum (Eq.~\ref{eq:deltaChi2}). The uncertainty is given by the standard deviation of the values from all the valid solutions. In addition, the validity of the fit of the photometric data to a specific spin/shape solution is quantified by the $\chi^2$. When using $3\sigma$ in order to determine the uncertainty range above the global minimum of the $\chi^2$, we can be sure that the model is correct in $99.7\%$, therefore, the main uncertainty in the reported values is mostly dependent on the standard deviation of the physical solutions. However, when using $1\sigma$ only, the validity of the spin/shape solutions is only correct in $\sim68\%$. Therefore, in the case of 3749, 7343 and 44612, where the uncertainty of the fit was determined by only $1\sigma$, the main uncertainty in the reported results is due to the fit of the spin/shape solution. However, the $3\sigma$-based uncertainty range are marked on the $\chi^2$ maps of the spin axes fits in order to show that the additional uncertainty does not significantly change the result. An exception is 7343, that its $3\sigma$-based uncertainty range has no significant meaning.

\subsection{Estimating the density of an asteroid pair}
\label{sec:EstimateDensity}
	Assuming the {\it asteroid pairs} were formed by a rotational fission event in their recent past, we can use the derived parameters of spin and shape as the input of a model and by that constrain their density values. Pravec and Harris (2007) and Pravec et al. (2010; Supplementary Information) presented a model based on the conservation of angular momentum. The model from the two papers is presented in appendix A, while keeping the symbols chosen by these authors for clarity.

	The model describes two bodies (the current primary and secondary pair members) laying one on top the other. The source of angular momentum for the two touching bodies is the spin of each object and their orbit around each other. Because the spin-orbit coupling is internal to the system, no energy is lost and the angular momentum is conserved. Photometric observations (Pravec et al. 2010) support this model, since a correlation was found between the current spin rate of the primary member $\omega_{1fin}$ to the mass ratio between the components $q$. An updated graph of the pairs' values and their correlation to the described model is displayed in Fig.~\ref{fig:pairs_q_P}. The correlation is described as follow:

\begin{figure}
\centerline{\includegraphics[width=8.5cm]{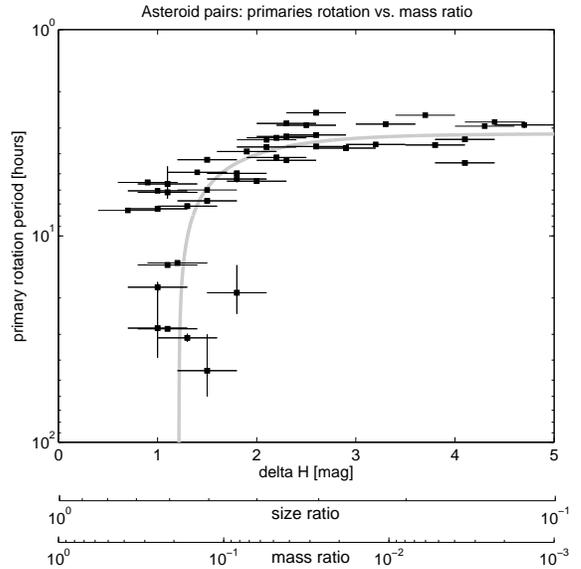}}
\caption{An updated version of the correlation between the rotation period of the primary to the size/mass ratio between the two components of each pair (black squares), originally published by Pravec et al. (2010). Additional data are from Slivan et al. 2008, Polishook 2011, Polishook et al. 2011 and Polishook 2014. In most cases, the error on the Y-axis is negligible. The rotational-fission model (grey line) is based on the assumption of a conservation of angular momentum. See the text for more details.
\label{fig:pairs_q_P}}
\end{figure}

\begin{equation}
\omega_{1fin}^2 = \omega_{1ini}^2 - \frac{20\pi\rho G\frac{a_1}{b_1}\frac{c_1}{b_1}}{3[1+(\frac{a_1}{b_1})^2]\frac{A_{ini}}{b_1}}q
\label{eq:MomConservShort_a}
\end{equation}

Where $\omega_{1ini}$ is the initial spin rate of the primary before the separation of the two objects, $\frac{a_1}{b_1}$ and $\frac{b_1}{c_1}$ are the ratios between the triaxial axes of the primary, $G$ is the gravitational constant, $\rho$ is the density, and the initial semi-major axis $A_{ini}$, normalized by the intermediate axis of the primary $b_1$, is of order unity for a contact binary.

	The initial spin of the primary $\omega_{1ini}$ can be formulated as follows:

\begin{eqnarray}
\omega_{1ini} = \frac{1}{(\frac{a_1^2}{b_1^2}+1)}[2\alpha_L(\frac{a_1}{b_1}\frac{c_1}{b_1})^{2/3}(1+q)^{5/3}\sqrt{\frac{4\pi}{3}G\rho} & \cr
-\frac{5q}{q+1}\sqrt{\frac{4\pi}{3}G\rho\frac{a_1}{b_1}\frac{c_1}{b_1}(1+q)\frac{A_{ini}}{b_1}(1-e^2)} \cr
-q(\frac{a_2^2}{b_2^2}+1)\omega_2]
\label{eq:MomConservShort_b}
\end{eqnarray}

Where $\omega_2$ is the supposedly unchanged spin rate of the secondary, $\frac{a_2}{b_2}$ is the ratios between the long triaxial axes of the secondary, and $e$ is the eccentricity of the secondary around the primary. $\alpha_L$ is the normalized total angular momentum of the system (see definition in Appendix A).

	I use this model in order to constrain the density of the asteroid pair. I applied it with the derived parameters from the lightcurve inversion method ($\omega_{1fin}=\frac{2\pi}{P_1}$, $a_1/b_1$, $b_1/c_1$). The mass ratio $q$ was obtained from the measured difference in magnitude $\Delta H$ as stated in the MPC website\footnote{$\frac{M_2}{M_1}=(\frac{D_2}{D_1})^3 \Rightarrow \frac{D_2}{D_1}=10^{(\frac{-\Delta H}{5})}$. This assumes that the two members have the same density and albedo, which is a reasonable assumption if the two objects were part of a single body.}. A value of 0.1 mag was chosen as the uncertainty of absolute magnitude obtained from the MPC. Combining the uncertainties of two asteroids to derive the formal uncertainty of $\Delta H$, I use a conservative value of 0.14 mag ($\sqrt(0.1^2+0.1^2)$) although the real uncertainty is probably smaller. $A_{ini}/b_1$ was set to two and zero eccentricity (e=0) was chosen. For $a_2/b_2$ I used measured values if existed or assume it to be 1 (spherical shape). If the rotation rate of the secondary ($\omega_2$) is unknown $P_2=2\pi / \omega_2$ was set to 2 hours, but due to the small size of the secondary members, this parameter is negligible. Because the rotation periods of the asteroids are known in high accuracy, the normalized total angular momentum, $\alpha_L$, is a function of the shape of the asteroids (parameters {\it a}, {\it b} and {\it c}). Pravec and Harris (2007) showed that for asteroids in zero tensile strength, like the asteroids studied here, there is an upper limit on $\alpha_L$ as a function of a/b (see their Fig. 1). For the derived values the upper limit is one. This is also the average of the measured values of $\alpha_L$ for binary asteroids (Pravec and Harris 2007). Therefore, in this model $\alpha_L$ was set to one.

\section{Results}
\label{sec:results}

Here I report spin and shape parameters of six asteroids in pairs that were derived from the lightcurve inversion method. The orbital and physical parameters of these asteroids are summarized in Table~\ref{tab:PairsParam}. As described above, for each object multiple possible solutions within the uncertainty range are derived, and all of them are treated as valid parameters. The average values for each parameter is presented and their standard deviation is used as the official error (summarized in Table~\ref{tab:Results} and~\ref{tab:PairsDensity}).

\begin{deluxetable*}{ccccccccccc}
\tablecolumns{11}
\tablewidth{0pt}
\tablecaption{Dynamical, physical and evolutional parameters}
\tablehead{
\colhead{Asteroid} &
\colhead{1/2} &
\colhead{a} &
\colhead{e} &
\colhead{i} &
\colhead{$H_v$} &
\colhead{D} &
\colhead{Taxonomy} &
\colhead{Partner} &
\colhead{$\Delta H$} &
\colhead{Age} \\
\colhead{} &
\colhead{} &
\colhead{[AU]} &
\colhead{} &
\colhead{[deg]} &
\colhead{[mag]} &
\colhead{[km]} &
\colhead{} &
\colhead{} &
\colhead{[mag]} &
\colhead{[kyrs]}
}
\startdata
2110	   & 1 & 2.198 & 0.178 & 1.130 & 13.2 & 6.5 & S		& 44612		& 2.3 & $>1600$		\\
3749	   & 1 & 2.237 & 0.109 & 5.382 & 13.1 & 6.8 & Sq		& 312497		& 4.4 &	$280^{-25}_{+45} $	\\
5026	   & 1 & 2.378 & 0.242 & 4.288 & 13.8 & 9.6 & Ch		& 2005WW113	& 4.0 & $18\pm1$		\\
6070	   & 1 & 2.388 & 0.210 & 3.131 & 13.7 & 5.2 & Sq		&  54827		& 1.6 & $17\pm0.5$	\\
7343	   & 1 & 2.193 & 0.139 & 3.959 & 13.9 & 4.7 & S		& 154634		& 3.0 & $>800$		\\
44612 & 2 & 2.199 & 0.178 & 1.121 & $15.51\pm0.02$ & 2.3 & Sq/Q & 2110 & 2.3 & $>1600$	
\enddata
\tablenotetext{}{Source of data:}
\tablenotetext{-}{Asteroids and their partners are from Vokrouhlick{\'y} \& Nesvorn{\'y} (2008), Pravec and Vokrouhlick{\'y} (2009) and Vokrouhlick{\'y} (2009).}
\tablenotetext{-}{Orbital elements, absolute magnitude $H_v$ and $\Delta H$ are from the MPC website. The absolute magnitude of 44612 was derived by this study. The official uncertainty choosen for the MPC's absolute magnitude is 0.1 mag.}
\tablenotetext{-}{Age and taxonomy of 2110, 3749, 5026, 6070 and 44612 are from Polishook et al. (2014), age of 7343 is from Pravec et al. (2010) and taxonomy is from Duddy et al. (2012).}
\tablenotetext{-}{Diameters were estimated from the absolute magnitude assuming an albedo value of 0.22 for S-complex and 0.058 for the Ch-type asteroid (Mainzer et al. 2011).}
\label{tab:PairsParam}
\end{deluxetable*}

\begin{deluxetable*}{ccccccccc}
\tablecolumns{9}
\tablewidth{0pt}
\tablecaption{Derived rotation state and shape model parameters}
\tablehead{
\colhead{Asteroid} &
\colhead{Apparitions} &
\colhead{Uncertainty $\sigma$} &
\colhead{Sidereal period} &
\colhead{Sense of rotation} &
\colhead{Axis $\lambda$} &
\colhead{Axis $\beta$} &
\colhead{a/b} &
\colhead{b/c} \\
\colhead{}          &
\colhead{}          &
\colhead{}          &
\colhead{[hour]}    &
\colhead{}          &
\colhead{[deg]} &
\colhead{[deg]} &
\colhead{}	&
\colhead{}
}
\startdata
2110	& 3 & 	3 &	$3.3447293\pm0.0000008$	& Retrograde			& ---			& $-70\pm10$	& $1.35\pm0.04$	& $1.1\pm0.1$	\\
3749	& 4 &	1 &	$2.80494\pm0.00007$		& $94\%$ Retrograde	& ---			& $-40\pm20$	& $1.11\pm0.04$	& $1.2\pm0.1$	\\
5026	& 4 & 	3 & 	$4.424087\pm0.000002$	& Prograde			& $354\pm8$		& $80\pm10$		& $1.68\pm0.08$	& $1.2\pm0.2$	\\
6070	& 4 &	3 & 	$4.273715\pm0.000003$	& Retrograde			& $110\pm20$	& $-60\pm10$	& $1.38\pm0.06$	& $1.2\pm0.2$	\\
		&    &  	   &							&					& $290\pm50$	&                      	&				&				\\
7343	& 3 &	1 & 	$3.7554\pm0.0003$		& $60\%$ Prograde		& $70\pm40$		& $50\pm20$		& $1.11\pm0.04$	& $1.3\pm0.2$	\\
		&    &    	   &							&					& $260\pm40$	& $-40\pm20$	&				&				\\
44612	& 3 & 	1 & 	$4.907068\pm0.000003$	& Retrograde			& $60\pm30$		& $-50\pm20$	& $1.11\pm0.03$	& $1.3\pm0.1$	\\
		&    &    	   &							&					& $240\pm30$	&				&				&				
\enddata
\label{tab:Results}
\end{deluxetable*}

\begin{deluxetable}{cc}
\tablecolumns{2}
\tablewidth{0pt}
\tablecaption{Derived density range from the rotational-fission model}
\tablehead{
\colhead{Asteroid} &
\colhead{density range} \\
\colhead{} &
\colhead{$gr ~ cm^{-3}$}
}
\startdata
2110	   & 1.60 to 2.43 \\
3749	   & 1.67 to 2.27 \\
5026	   & 0.97 to 1.80 \\
6070	   & 1.45 to 4.40 \\
7343	   & 1.01 to 1.75 
\enddata
\label{tab:PairsDensity}
\end{deluxetable}

\subsection{(2110) Moore-Sitterly and (44612) 1999RP27}
\label{sec:moore-sitterly}

	(2110) {\it Moore-Sitterly} and its secondary member (44612) {\it 1999RP27} are among the ``oldest" known pairs with a proposed separation time of more than $\sim1.6$ Myrs ago (Polishook et al. 2014). The drifting velocity between the two objects in the proper element space ($3.36 m/s$; Pravec \& Vokrouhlick{\'y} 2009) is low as expected from an object splitted by rotational-fission. The S-complex near-IR reflectance of both members match each other supporting a common origin (Polishook et al. 2014). Both of their rotation periods were measured, and the period of 2110 fits with the rotational-fission model suggested by Pravec et al. (2010). These measurements support the notion that 2110 and 44612 are truly an asteroid pair that was formed by the rotational-fission mechanism.

\begin{figure}
\centerline{\includegraphics[width=8.5cm]{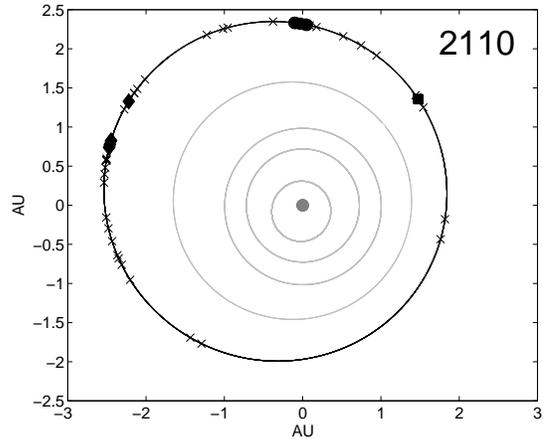}}
\caption{The locations on the orbit of (2110) {\it Moore-Sitterly} where it was observed for this study. 2110 was observed on December 2008 to January 2009 (circles), October 2011 (squares) and February 2013 (diamonds). Data taken by large-scale surveys (x) were collected between November 5, 1999 to May 29, 2013.
\label{fig:2110_obs}}
\end{figure}

\begin{figure}
\centerline{\includegraphics[width=8.5cm]{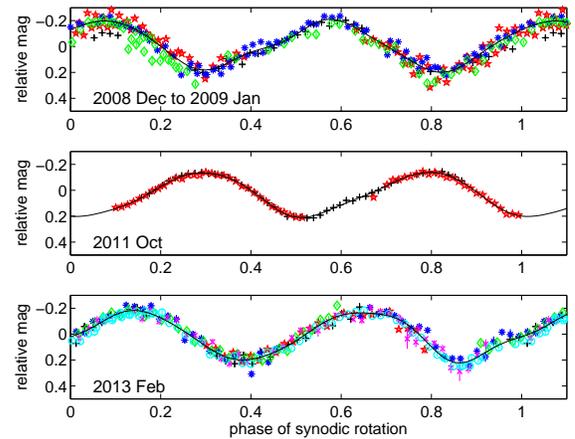}}
\caption{Folded lightcurves of (2110) {\it Moore-Sitterly} from three apparitions. The epoch is $T_0=2454820 JD$. Different markers and colors represent photometric data from different observing nights (see Table~\ref{tab:ObsCircum1}). The x-axis is the synodic rotation phase. The synodic rotation period is $3.3447\pm0.0003$ hours. The difference in the amplitude is due to the different aspect angle and phase angle at the three apparitions (see Table~\ref{tab:ObsCircum1}).
\label{fig:2110_LC}}
\end{figure}

\begin{figure}
\centerline{\includegraphics[width=8.5cm]{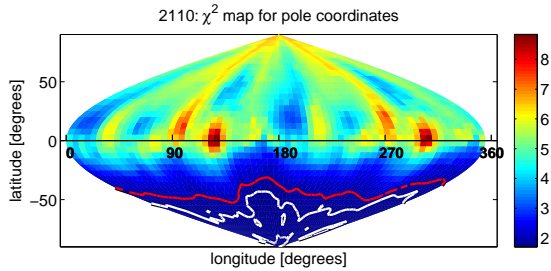}}
\caption{The $\chi^2$ values for all spin axis solutions on a longitude-latitude plane for (2110) {\it Moore-Sitterly}. The uncertainty of the fit corresponding to $1\sigma$ (white solid line) and $3\sigma$ (red line) above the global minimum clearly demonstrate the retrograde sense of rotation of 2110. The globally best-fit solution has $\chi^2$ = 1.70.
\label{fig:2110_axis}}
\end{figure}

\begin{figure}
\centerline{\includegraphics[width=8.5cm]{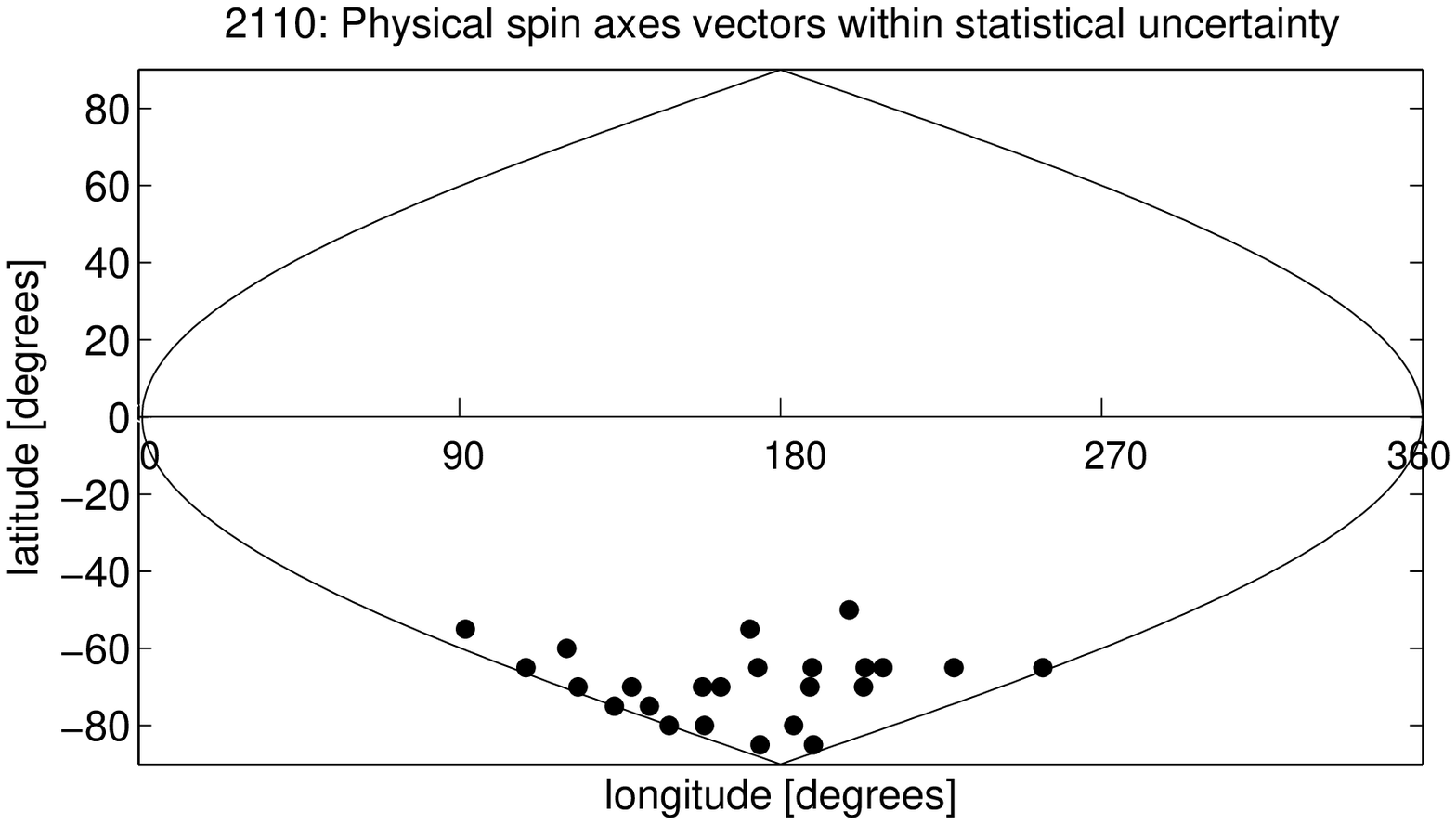}}
\caption{Possible coordinates of spin axes that have feasible physical parameters for (2110) {\it Moore-Sitterly}. These constrain the latitude of the spin axis to $-70\pm10$ degrees, where the uncertainty is the standard deviation of the latitude of all feasibile axes.
\label{fig:2110_feasAxis}}
\end{figure}

\begin{figure}
\centerline{\includegraphics[width=8.5cm]{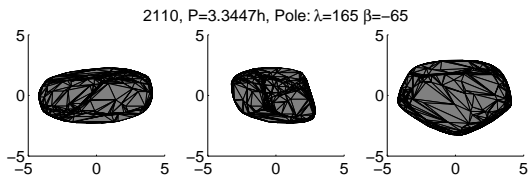}}
\caption{A typical shape model for (2110) {\it Moore-Sitterly} from the lightcurve inversion analysis. The three views are from equatorial level (left and center) and pole-on (right).
\label{fig:2110_shape1}}
\end{figure}

	Photometric observations obtained on 14 nights and 3 apparitions at the Wise Observatory (Fig.~\ref{fig:2110_obs}) were used for the analysis of 2110. To this dataset 39 sets of sparse photometric data, collected by the Catalina Sky Survey, were added. The lightcurves have high SNR and low photometric errors (0.01 to 0.02 mag; Fig.~\ref{fig:2110_LC}). The analysis results with 24 plausible spin axis solutions within the uncertainty range that corresponds to $3\sigma$ above the global minimum and with physical feasibility (Fig.~\ref{fig:2110_axis}-~\ref{fig:2110_feasAxis}). All solutions present a retrograde sense of rotation with a mean latitude of $-70^o\pm10^o$. The longitude of the spin axis is not constrained. The shape models that match to the spin axis solutions have all similar northern hemisphere solution, with a shape of an irregular pentagon (e.g. Fig~\ref{fig:2110_shape1}). The southern hemisphere, though, is less constrained and different shapes were derived.

\begin{figure}
\centerline{\includegraphics[width=8.5cm]{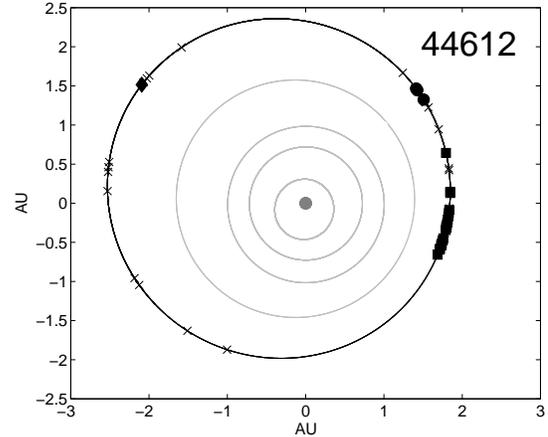}}
\caption{The locations on the orbit of (44612) {\it 1999RP27} where it was observed for this study. 44612 was observed on October 2009 (circles), August to October 2012 (squares) and February 2014 (diamonds). Data taken by large-scale surveys (x) was observed between August 22, 1999 to February 4, 2014.
\label{fig:44612_obs}}
\end{figure}

\begin{figure}
\centerline{\includegraphics[width=8.5cm]{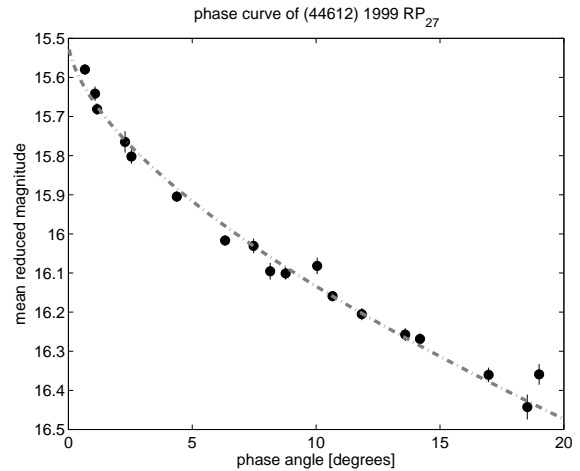}}
\caption{A phase curve of (44612) {\it 1999RP27} constructed using data obtained on the apparition of 2012. A {\it H-G} model fitted to the phase curve (dash line) result with $H_V=15.51\pm0.02$ and a {\it G} slope of $0.18\pm0.02$.
\label{fig:44612_PC}}
\end{figure}

\begin{figure}
\centerline{\includegraphics[width=8.5cm]{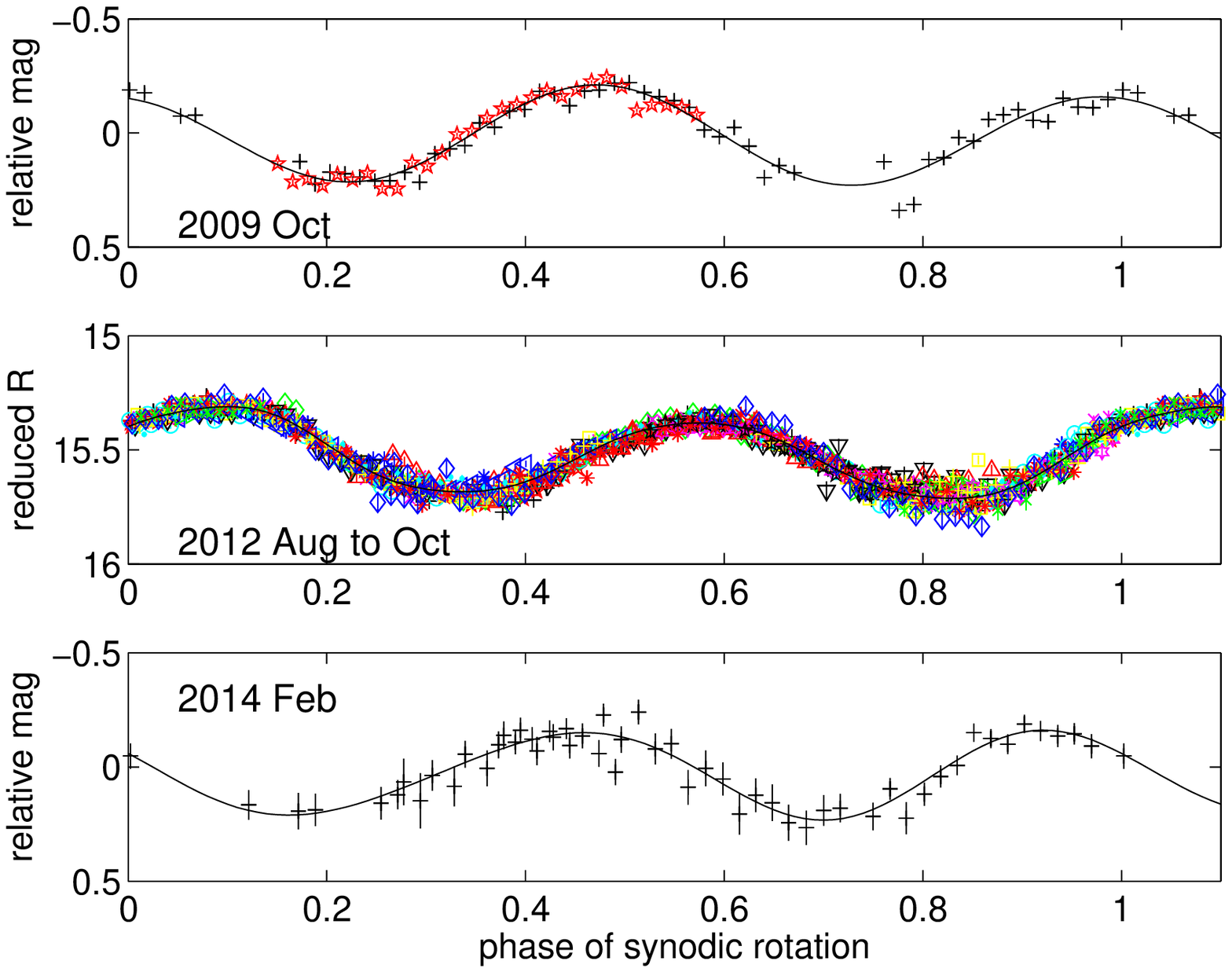}}
\caption{Folded lightcurves of (44612) {\it 1999RP27} from three apparitions. The epoch is $T_0=2455127 JD$. Different markers and colors represent photometric data from different observing nights (see Table~\ref{tab:ObsCircum1}). The x-axis is the synodic rotation phase. The synodic rotation period is $4.9066\pm0.0003$ hours.
\label{fig:44612_LC}}
\end{figure}

	The secondary member of this pair, (44612) {\it 1999RP27}, was observed from the Wise Observatory on three appariations (Fig.~\ref{fig:44612_obs}). On the second apparition the asteroid was regularly sampled on a large range of phase angles. The photometric data were calibrated to the standard level and the phase curve (Fig.~\ref{fig:44612_PC}) was fitted using the {\it H-G} system (Bowell et al. 1989) in order to derive the absolute magnitude of 44612, and a more accurate value of $\Delta H$ of the pair. A visible magnitude of $H_V=15.51\pm0.02$ (equal to the MPC's value) and a {\it G} slope of $0.18\pm0.02$ were derived. The folded lightcurves are presented on Fig.~\ref{fig:44612_LC}. 26 sets of measurements collected by the Catalina Sky Survey were used to expand the dataset.

\begin{figure}
\centerline{\includegraphics[width=8.5cm]{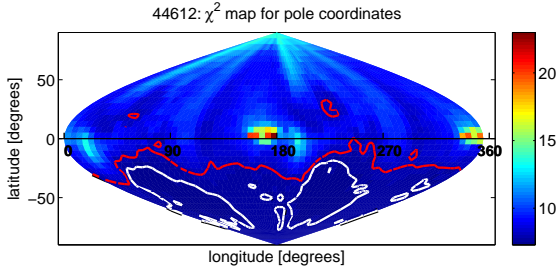}}
\caption{The $\chi^2$ values for all spin axis solutions on a longitude-latitude plane for (44612) {\it 1999RP27}. The uncertainty of the fit corresponding to $1\sigma$ (white solid line) and $3\sigma$ (red line) above the global minimum demonstrate the retrograde sense of rotation of 44612.
\label{fig:44612_axis}}
\end{figure}

\begin{figure}
\centerline{\includegraphics[width=8.5cm]{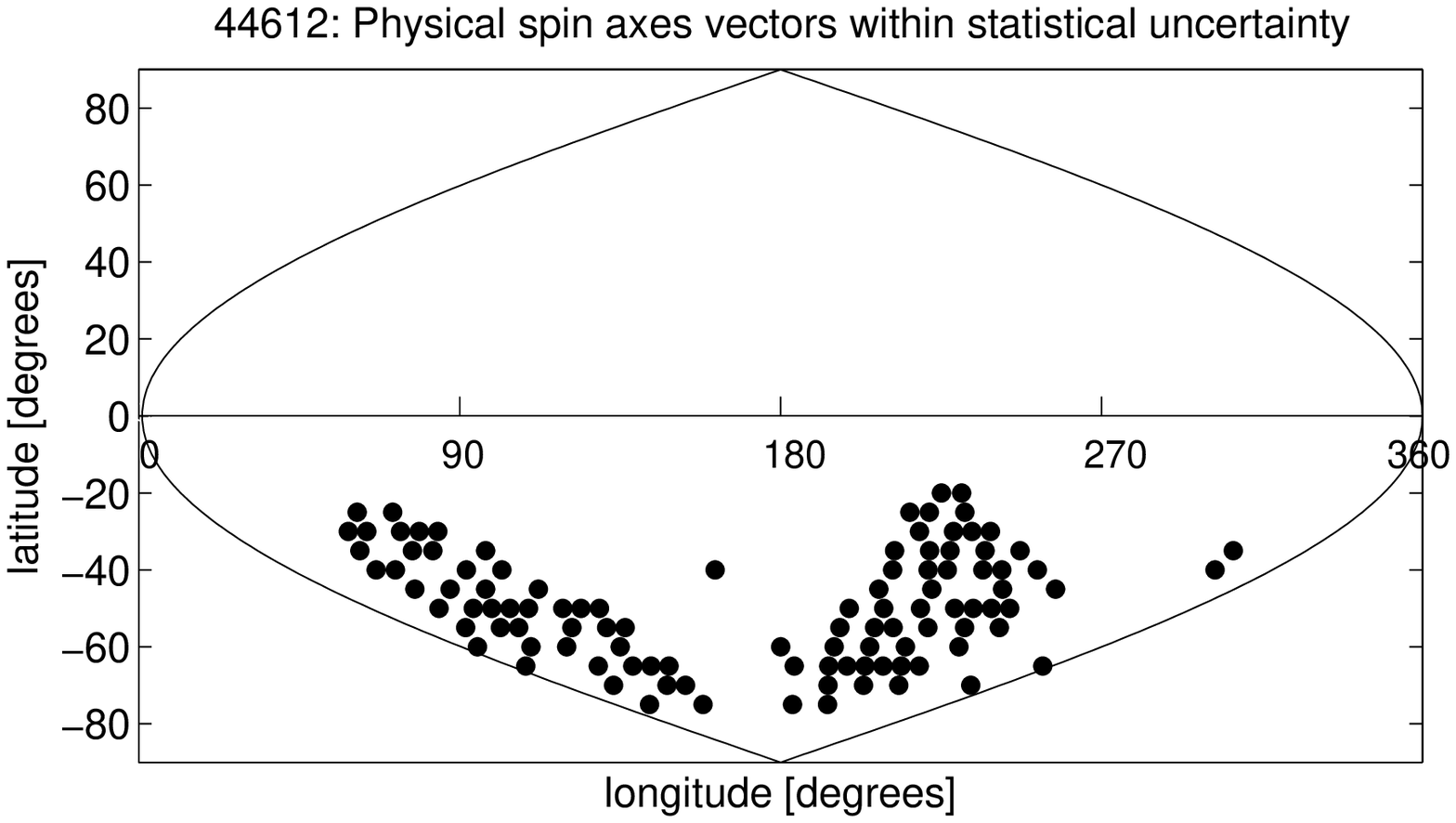}}
\caption{Possible coordinates of spin axes that have feasible physical parameters for (44612) {\it 1999RP27}. These constrain the latititude of the spin axis of this asteroid to $-50\pm20$ degrees, where the uncertainty is the standard deviation of the latitude of all feasibile axes.
\label{fig:44612_feasAxis}}
\end{figure}

\begin{figure}
\centerline{\includegraphics[width=8.5cm]{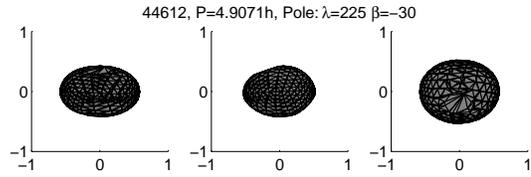}}
\caption{A typical shape model for (44612) {\it 1999RP27} from the lightcurve inversion analysis. The three views are from equatorial level (left and center) and pole-on (right).
\label{fig:44612_shape1}}
\end{figure}

	The smaller photometric data of 44612 put less constraints on the models compare to its companion, thus more spin axes solutions fall within the uncertainty range as determined by the $\chi^2$ results (Fig.~\ref{fig:44612_axis}). However, all have a retrograde sense of orbit within an uncertainty range that corresponds to $1\sigma$ above the global minimum and with physical feasibility (Fig.~\ref{fig:44612_axis}$-$~\ref{fig:44612_feasAxis}). This means that the two members of this pair (2110 and 44612) have both the same sense of rotation. The longitude of the two spin axes are not constrained enough, but the latitude falls within the same uncertainty range ($\beta_{2110}=-70\pm10$, $\beta_{44612}=-50\pm20$). The shape models of 44612 that match to the spin axis solutions are all rounded with no pointy edges (e.g., Fig.~\ref{fig:44612_shape1}). It is premature to interpret this shape as the end result of the rotational-fission mechanism for the ejected secondary component, since this shape might also be the result of the smaller photometric coverage of the asteroid on its orbit.

By applying the separation model (Eq.~\ref{eq:MomConservShort_a}$-$~\ref{eq:MomConservShort_b}) on the derived axis and shape parameters the density of 2110 is constrained to $1.6$ to $2.4~gr~cm^{-3}$. This density range is low compared to the density of ordinary chondrites that S-complex asteroids are made of ($\sim3.3 ~ gr ~ cm^{-3}$; Carry 2012). This density range is centered on the density of (25143) {\it Itokawa}, a rubble pile near-Earth asteroid that was visited and measured by the spacecraft {\it Hayabusa} (Fujiwara et al. 2006). This supports the notion that 2110 is a rubble pile asteroid that could have split due to the rotational-fission mechanism.

\begin{figure}
\centerline{\includegraphics[width=8.5cm]{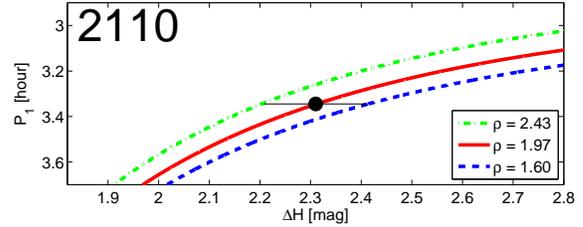}}
\caption{The measured rotation period of (2110) {\it Moore-Sitterly} and its size/mass ratio with its secondary (black circle) compared to the values obtained by the model of angular momentum conservation (lines). The three lines represent the mean and two extreme cases with the stated density ($\rho$) values. These include the extreme cases of the triaxial axes of the shape model as derived from our analysis. The error on the Y-axis is negligible. The density range of (1.60 to 2.43 $gr ~ cm^{-3}$) for an S-complex object supports the notion that 2110 is a rubble pile asteroid that could have split due to rotational-fission.
\label{fig:2110_dens}}
\end{figure}

\subsection{(3749) Balam}
\label{sec:balam}

	(3749) {\it Balam} is a unique system of asteroids. It has a nearby satellite (at a separation of 20 km from the main body), that was discovered by light attenuation within 3749's lightcurves due to mutual events (Marchis et al., 2008a); a distant satellite (at $\sim290$ km from the main body), revealed in direct images on the Gemini-North telescope (Merline et al. 2002; the satellite's orbit was measured by Marchis et al. 2008b); and a secondary ``{\it pair}", designated as 312497, found by numerical integrations (Vokrouhlick{\'y} 2009) to have detached from the main body $280^{+45}_{-25}$ kyr ago (Polishook et al. 2014). It is unknown if all these objects were formed on the same fission event.

\begin{figure}
\centerline{\includegraphics[width=8.5cm]{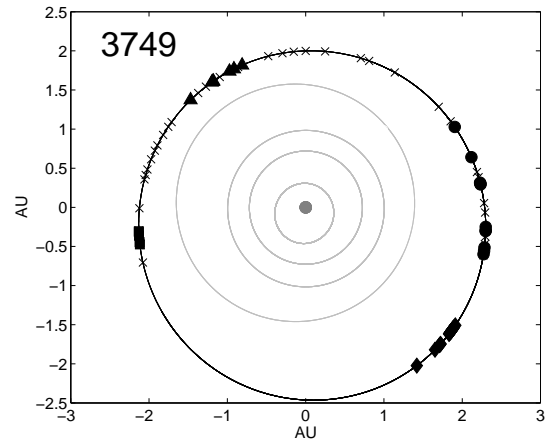}}
\caption{The locations on the orbit of (3749) {\it Balam} where it was observed for this study. 3749 was observed on July to December 2007 (circles), March 2009 (squares), June to August 2010 (diamonds) and December 2011 to February 2012 (triangulars). Data taken by large-scale surveys (x) were collected between December 15, 1999 to November 26, 2013.
\label{fig:3749_obs}}
\end{figure}

\begin{figure}
\centerline{\includegraphics[width=8.5cm]{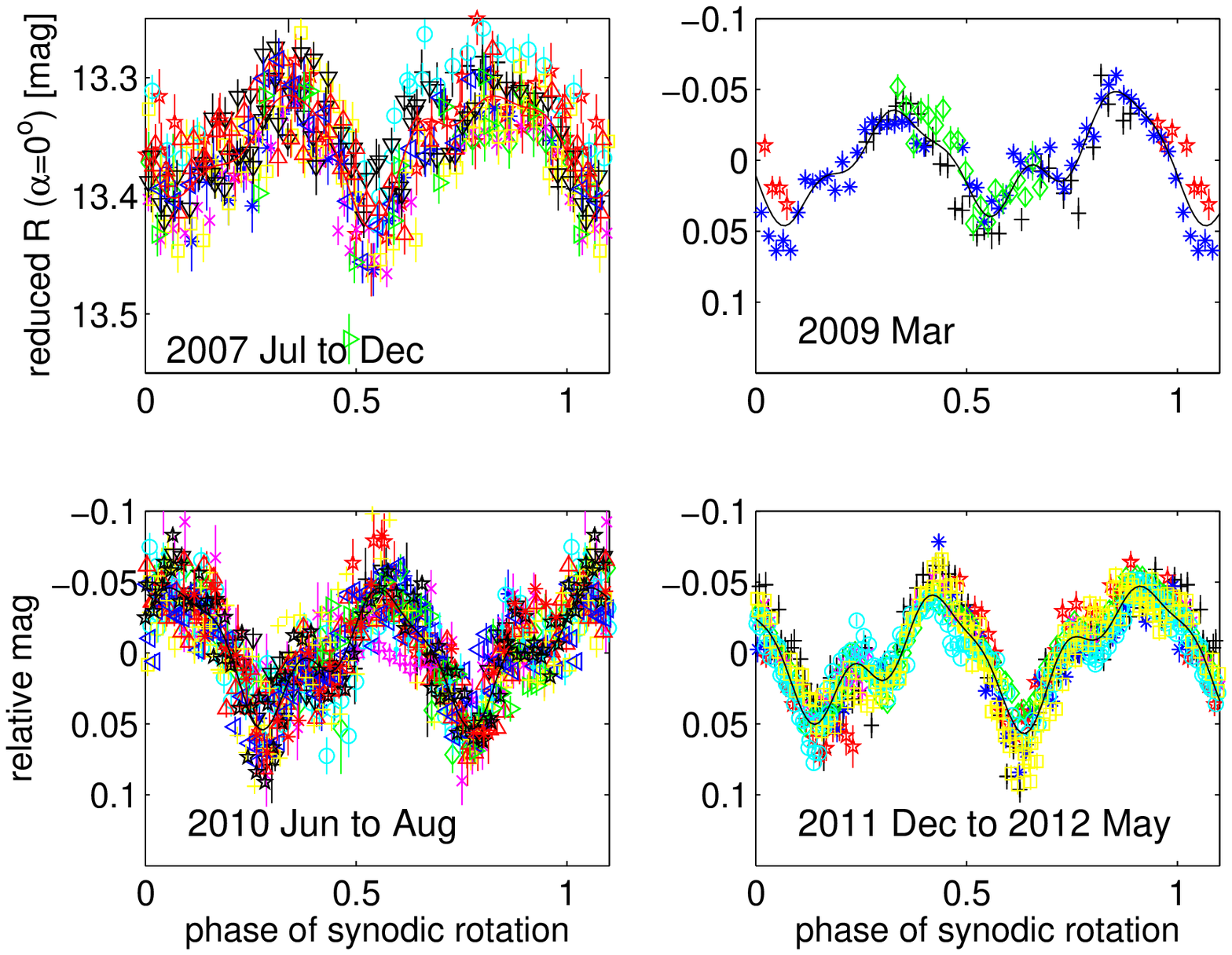}}
\caption{Folded lightcurves of (3749) {\it Balam} from four apparitions. The epoch is $T_0=2454297 JD$. Different markers and colors represent photometric data from different observing nights (see Table~\ref{tab:ObsCircum1}). The x-axis is the synodic rotation phase. The synodic rotation period is $2.8049\pm0.0002$ hours. The difference in the amplitude is due to the different aspect angle and phase angle at the four apparitions (see Table~\ref{tab:ObsCircum1}).
\label{fig:3749_LC}}
\end{figure}

\begin{figure}
\centerline{\includegraphics[width=8.5cm]{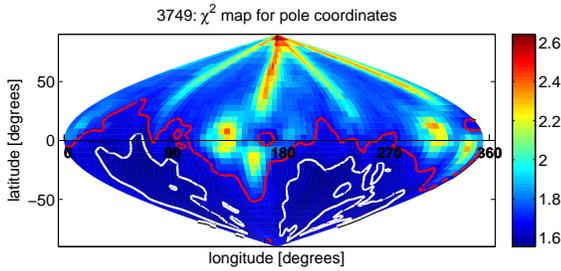}}
\caption{The $\chi^2$ values for spin axis solutions with a period of 2.8049 hours on a longitude-latitude plane for (3749) {\it Balam}. Over $90\%$ of the possible solutions originated from the analysis using this sidereal period. The rest of the solutions are obtained when two other sidereal periods are used and the results have $\chi^2$ within the uncertainty range plotted here. The uncertainty of the fit corresponding to $1\sigma$ (white solid line) and $3\sigma$ (red line) above the global minimum. The globally best-fit solution has $\chi^2$ = 1.55.
\label{fig:3749_axis}}
\end{figure}

\begin{figure}
\centerline{\includegraphics[width=8.5cm]{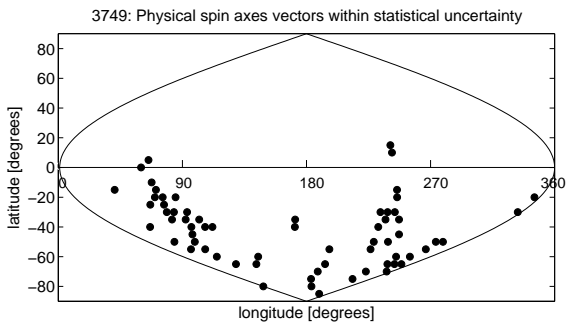}}
\caption{Possible coordinates of spin axes that have feasible physical parameters for (3749) {\it Balam}. $\sim95\%$ of the possible spin axes has retrograde rotations.
\label{fig:3749_feasAxis}}
\end{figure}

\begin{figure}
\centerline{\includegraphics[width=8.5cm]{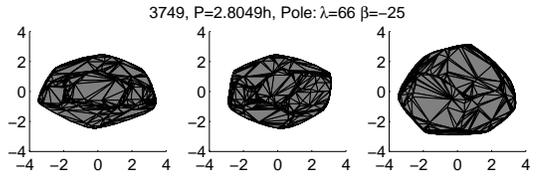}}
\caption{A typical shape model for (3749) {\it Balam} from the lightcurve inversion analysis. The three views are from equatorial level (left and center) and pole-on (right).
\label{fig:3749_shape1}}
\end{figure}

	3749 was observed from the Wise Observatory on 38 nights during four apparitions (2007 to 2012; Fig.~\ref{fig:3749_obs}). Additional usable observations were impossible to obtain in 2013 because the galaxy served as a dense background. Measurements from 35 additional nights collected by the Catalina Sky Survey were added to the photometric dataset. Mutual events were removed (Fig.~\ref{fig:3749_LC}) to allow a continuous analysis. Using a less restrictive $1\sigma$ above the global minimum (Fig.~\ref{fig:3749_axis}), the analysis finds 65 plausible spin axes (Fig.~\ref{fig:3749_feasAxis}), $\sim95\%$ of them have a retrograde sense of rotation, while the other $5\%$ points to the equator. A $3\sigma$-based uncertainty range adds more plausible axes but these cross the equator only slightly; therefore, 3749 most probably rotates in a retrograde rotation. The derived shape models of 3749 present repeated pattern of non-elongated body with low {\it a/b} ratio (e.g. Fig.~\ref{fig:3749_shape1}).
	
\begin{figure}
\centerline{\includegraphics[width=8.5cm]{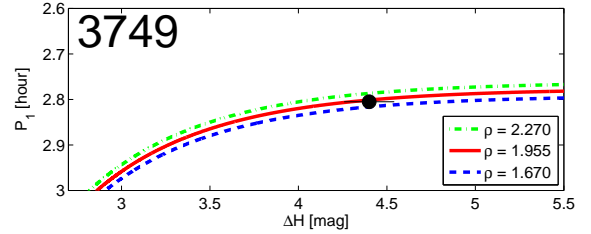}}
\caption{The measured rotation period of (3749) {\it Balam} and its size/mass ratio with its secondary (black circle) compared to the values obtained by the model of angular momentum conservation (lines). The three lines represent the mean and two extreme cases with the stated density ($\rho$) values. These include the extreme cases of the triaxial axes of the shape model as derived from our analysis. The error on the Y-axis is negligible. The density range of (1.67 to 2.27 $gr ~ cm^{-3}$) for an S-complex object supports the notion that 3749 is a rubble pile asteroid that could have split due to rotational-fission.
\label{fig:3749_dens}}
\end{figure}

	Fitting the rotational-fission model to the measured $\Delta H$ and spin rate, the density is constrained to 1.7 to 2.3  $gr ~ cm^{-3}$ (Fig.~\ref{fig:3749_dens}). The low uncertainty range of the density is mainly due to the small size of the secondary member compared to the size of 3749, since in this regime the model is highly dependent on the rotation period that has much smaller uncertainty than on the mass ratio of the pair. Similar to the case of 2110, this range matches the expected density of a ``rubble pile" body that is composed of ordinary chondrites (S-complex taxonomy) which is the composition of 3749 (Polishook et al. 2011, 2014). However, our simplistic model does not take into account the complexity of 3749 system with its two satellites. If the secondary pair, 312497, was ejected from 3749 with the currently two satellites than our model does not consider additional angular momentum in the system that is relevant to the rotational-fission mechanism; if 312497 was formed separately than the two satellites, our model might still be valid since the angular momentum originated from the rotational and orbital periods of the satellites cancel out and does not alter the result of the model.	

\subsection{(5026) Martes}
\label{sec:martes}
	(5026) {\it Martes} and its secondary member {\it 2005WW113} are one of the youngest known asteroid pairs with a proposed separation time of about $\sim18,000$ years ago as indicated by backward dynamical integration (Vokrouhlick{\'y} \& Nesvorn{\'y} 2008, Pravec et al. 2010, Polishook et al. 2014). This Ch-type asteroid (Polishook et al. 2014) is also unique in the sense it is much larger than its secondary member - with a $\Delta H=4$, their $D2/D1=0.16$ and mass ratio of less than $1\%$.

\begin{figure}
\centerline{\includegraphics[width=8.5cm]{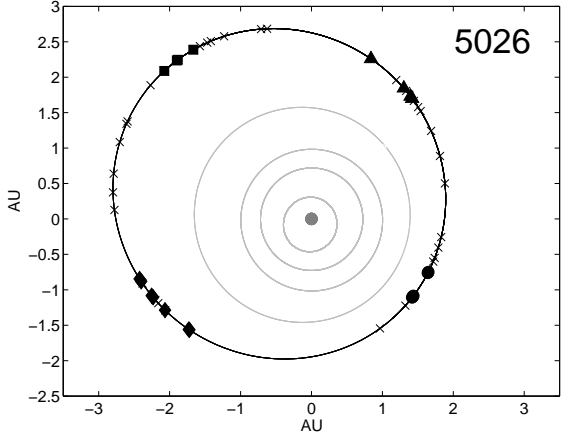}}
\caption{The locations on the orbit of (5026) {\it Martes} where it was observed for this study. 5026 was observed on July to August 2009 (circles), March 2011 (squares), March to June 2012 (diamonds) and October 2013 to January 2014 (triangulars). Data taken by large-scale surveys (x) were collected between December 15, 1999 to November 26, 2013.
\label{fig:5026_obs}}
\end{figure}

\begin{figure}
\centerline{\includegraphics[width=8.5cm]{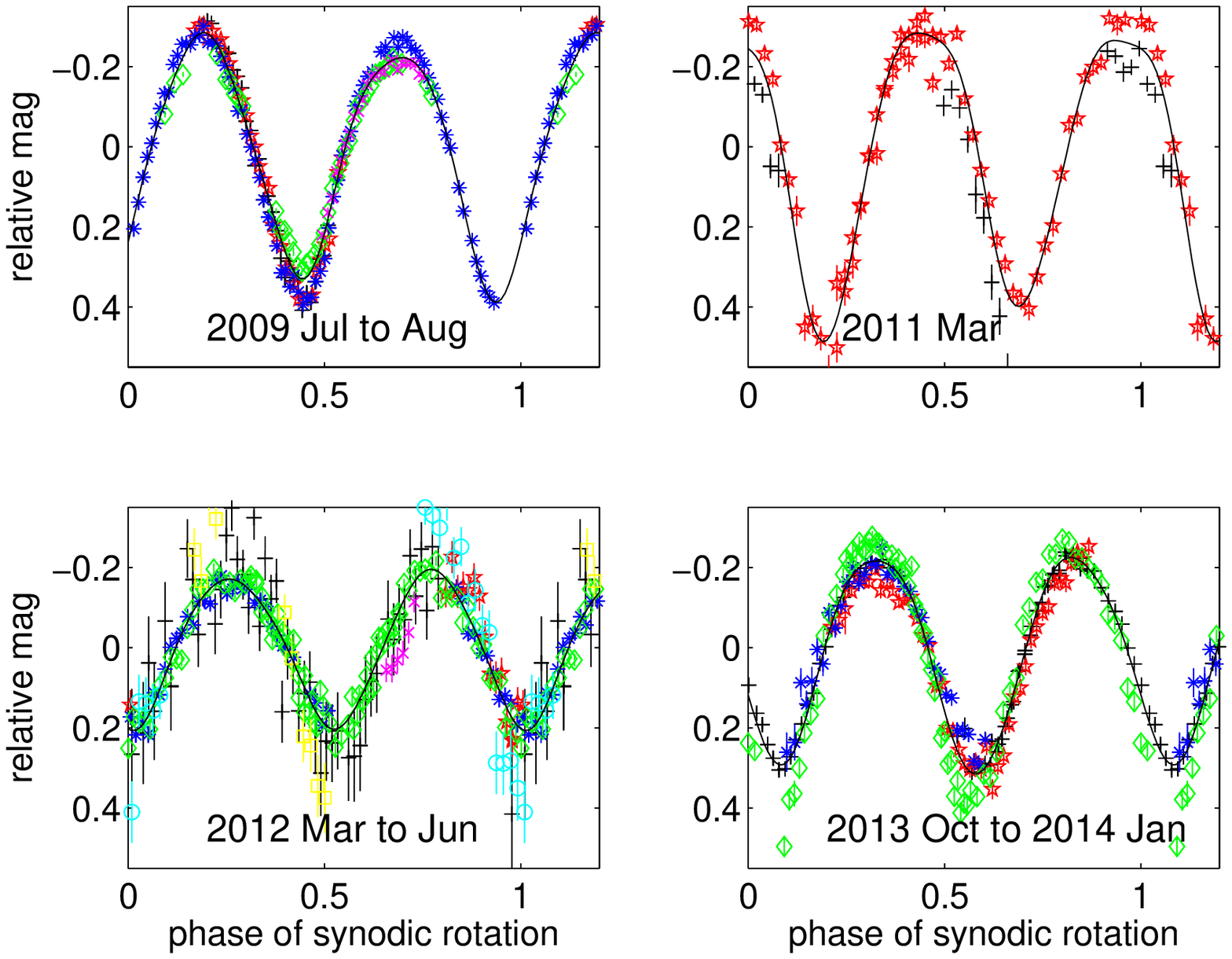}}
\caption{Folded lightcurves of (5026) {\it Martes} from four apparitions. The epoch is $T_0=2455028 JD$. Different markers and colors represent photometric data from different observing nights (see Table~\ref{tab:ObsCircum1}). The x-axis is the synodic rotation phase. The synodic rotation period is $4.4243\pm0.0001$ hours.
\label{fig:5026_LC}}
\end{figure}

\begin{figure}
\centerline{\includegraphics[width=8.5cm]{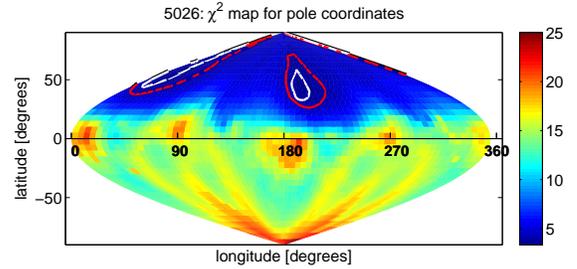}}
\caption{The $\chi^2$ values for all spin axis solutions on a longitude-latitude plane for (5026) {\it Martes}. The uncertainty of the fit corresponding to $1\sigma$ (white solid line) and $3\sigma$ (red line) above the global minimum, clearly demonstrate the prograde sense of rotation of 5026. The globally best-fit solution has $\chi^2$ = 3.60.
\label{fig:5026_axis}}
\end{figure}

\begin{figure}
\centerline{\includegraphics[width=8.5cm]{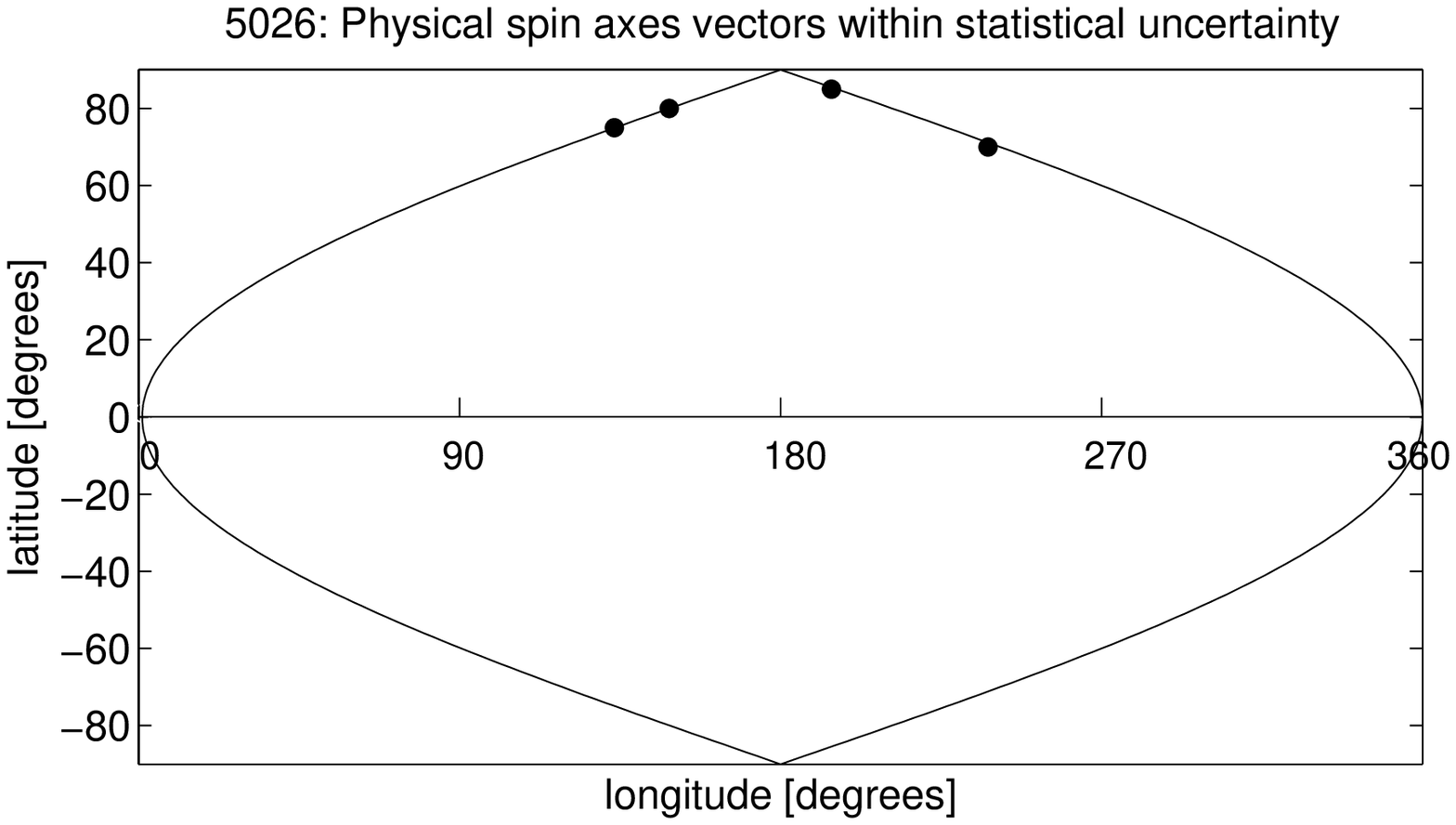}}
\caption{Possible coordinates of spin axes that have feasible physical parameters for (5026) {\it Martes}. These constrain the latititude of the spin axis to $80\pm10$ degrees, where the uncertainty is the standard deviation of the latitude of all feasibile axes.
\label{fig:5026_feasAxis}}
\end{figure}

\begin{figure}
\centerline{\includegraphics[width=8.5cm]{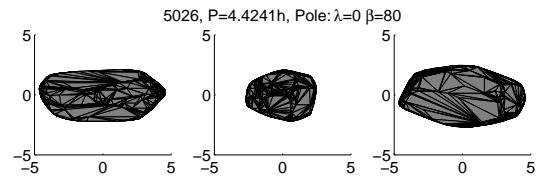}}
\caption{A typical shape model for (5026) {\it Martes} from the lightcurve inversion analysis. All the four possible shape models derived are generally similar. The three views are from equatorial level (left and center) and pole-on (right).
\label{fig:5026_shape}}
\end{figure}

	I observed 5026 at Wise Observatory on 18 nights during 4 apparitions running from 2009 to 2014 (Fig.~\ref{fig:5026_obs}). The photometric set was increased with measurements from 36 nights collected by the Catalina Sky Survey between 1999 and 2013. The relatively large set of data, with high S/N (Fig.~\ref{fig:5026_LC}), taken on equally sampled geometries (Fig.~\ref{fig:5026_obs}), results in a well-defined coordinates for the spin axis and shape of 5026 (Fig.~\ref{fig:5026_axis}). Even though some spin axis solutions with mid-latitude poles are within the uncertainty range, the physical feasibility test rejects them as unphysical (Fig.~\ref{fig:5026_feasAxis}). 5026 is a relatively elongated object ($a/b=1.68\pm0.08$; Fig.~\ref{fig:5026_shape}) with low obliquity ($\beta_{axis}=80\pm10$; $\epsilon=14\pm9$ degrees).
	
\begin{figure}
\centerline{\includegraphics[width=8.5cm]{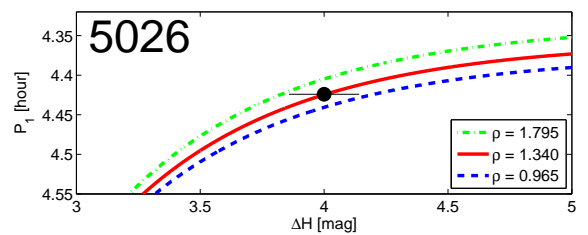}}
\caption{The measured rotation period of (5026) {\it Martes} and its size/mass ratio with its secondary (black circle) compared to the values obtained by the model of angular momentum conservation (lines). The three lines represent the mean and two extreme cases with the stated density ($\rho$) values. These include the extreme cases of the triaxial axes of the shape model as derived from our analysis. The error on the Y-axis is negligible. The density range of (0.965 to 1.795 $gr ~ cm^{-3}$) for a C-complex object supports the notion that 5026 is a rubble pile asteroid that could have split due to rotational-fission.
\label{fig:5026_dens}}
\end{figure}

	Fitting the rotational-fission model to the measured $\Delta H$ and spin rate, the density is constrained to 1.0 to 1.8 $gr ~ cm^{-3}$ (Fig.~\ref{fig:5026_dens}). This low range of density values, compared to 2110 and 3749, is consistent with a rubble pile structured body of the Ch-type taxonomy that represents a composition of low density carbonaceous chondrites ($\sim2.6 ~ gr ~ cm^{-3}$; DeMeo et al. 2009, Carry 2012). 
	
\subsection{(6070) Rheinland}
\label{sec:rheinland}
	(6070) {\it Rheinland} and its secondary member (54827) {\it 2001NQ8} are one of the youngest known asteroid pairs with a proposed separation time of $\sim17,000$ years ago as indicated by backward dynamical integration (Vokrouhlick{\'y} \& Nesvorn{\'y} 2008). Both components present reflectance spectrum typical of an ordinary chondrite (S-complex) supporting a common origin (Polishook et al. 2014). Furthermore, the visible and near-IR reflectance spectrum of 54827 has a low spectral slope and a deep absorption band at $1\mu m$ (Polishook et al. 2014), which are known parameters for recently exposed ordinary chondrite that was not exposed to space weathering agents (Binzel et al. 1996, 2010, Clark et al. 2002, Brunetto et al. 2006). Therefore, this is a supporting measurement for the young age of the system. A statistical model of the separation of this pair allowed Vokrouhlick{\'y} \& Nesvorn{\'y} (2009) to predict that 6070 has a retrograde sense of rotation. This claim was tested and confirmed by photometric observations that constrained the latitude of the spin axis to be less than $\sim-50^0$ (Vokrouhlick{\'y} et al. 2011).

\begin{figure}
\centerline{\includegraphics[width=8.5cm]{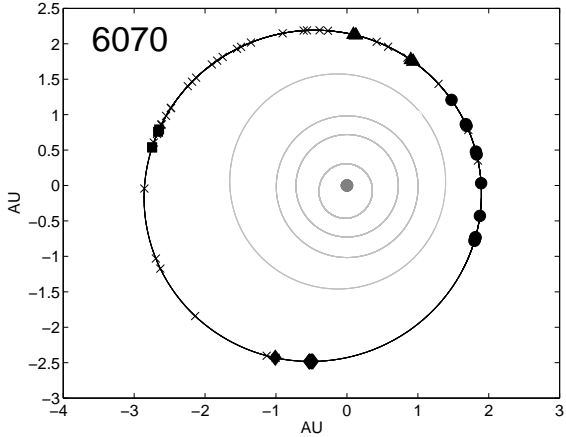}}
\caption{The locations on the orbit of (6070) {\it Rheinland} where it was observed for this study. 6070 was observed on July to December 2009 (circles), March 2011 (squares), May to July 2012 (diamonds) and October 2013 to January 2014 (triangulars). Data taken by large-scale surveys (x) were collected between September 21, 1998 to December 10, 2013.
\label{fig:6070_obs}}
\end{figure}

\begin{figure}
\centerline{\includegraphics[width=8.5cm]{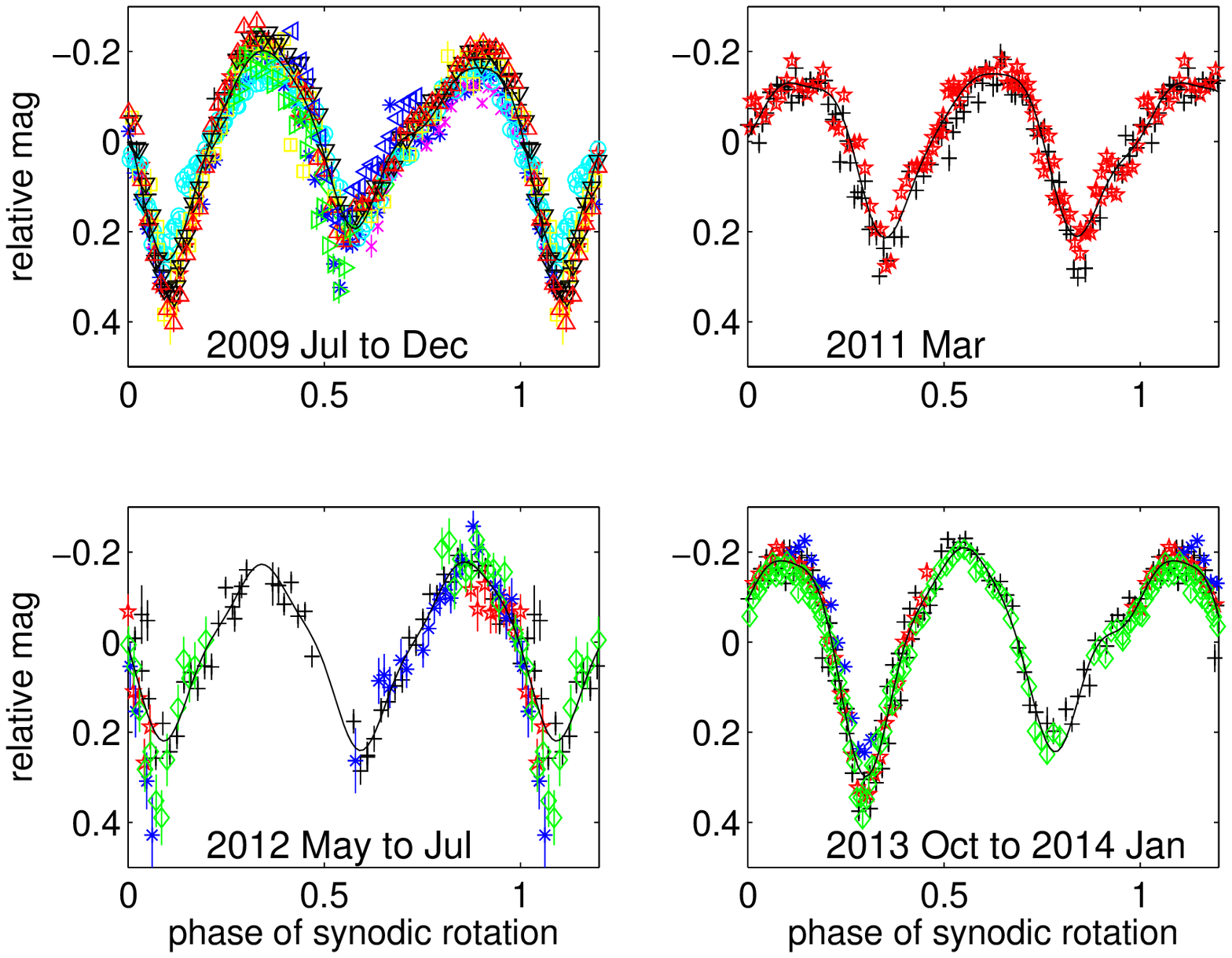}}
\caption{Folded lightcurves of (6070) {\it Rheinland} from four apparitions. The epoch is $T_0=2455034 JD$. Different markers and colors represent photometric data from different observing nights (see Table~\ref{tab:ObsCircum1}). The x-axis is the synodic rotation phase. The synodic rotation period is $4.2733\pm0.0001$ hours.
\label{fig:6070_LC}}
\end{figure}

\begin{figure}
\centerline{\includegraphics[width=8.5cm]{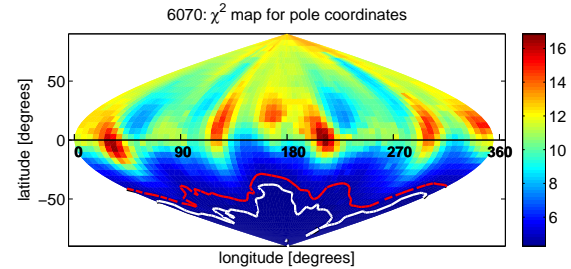}}
\caption{The $\chi^2$ values for all spin axis solutions on a longitude-latitude plane for (6070) {\it Rheinland}. The uncertainty of the fit corresponding to $1\sigma$ (white solid line) and $3\sigma$ (red line) above the global minimum clearly demonstrate the retrograde sense of rotation of 6070. The globally best-fit solution has $\chi^2$ = 4.29.
\label{fig:6070_axis}}
\end{figure}

\begin{figure}
\centerline{\includegraphics[width=8.5cm]{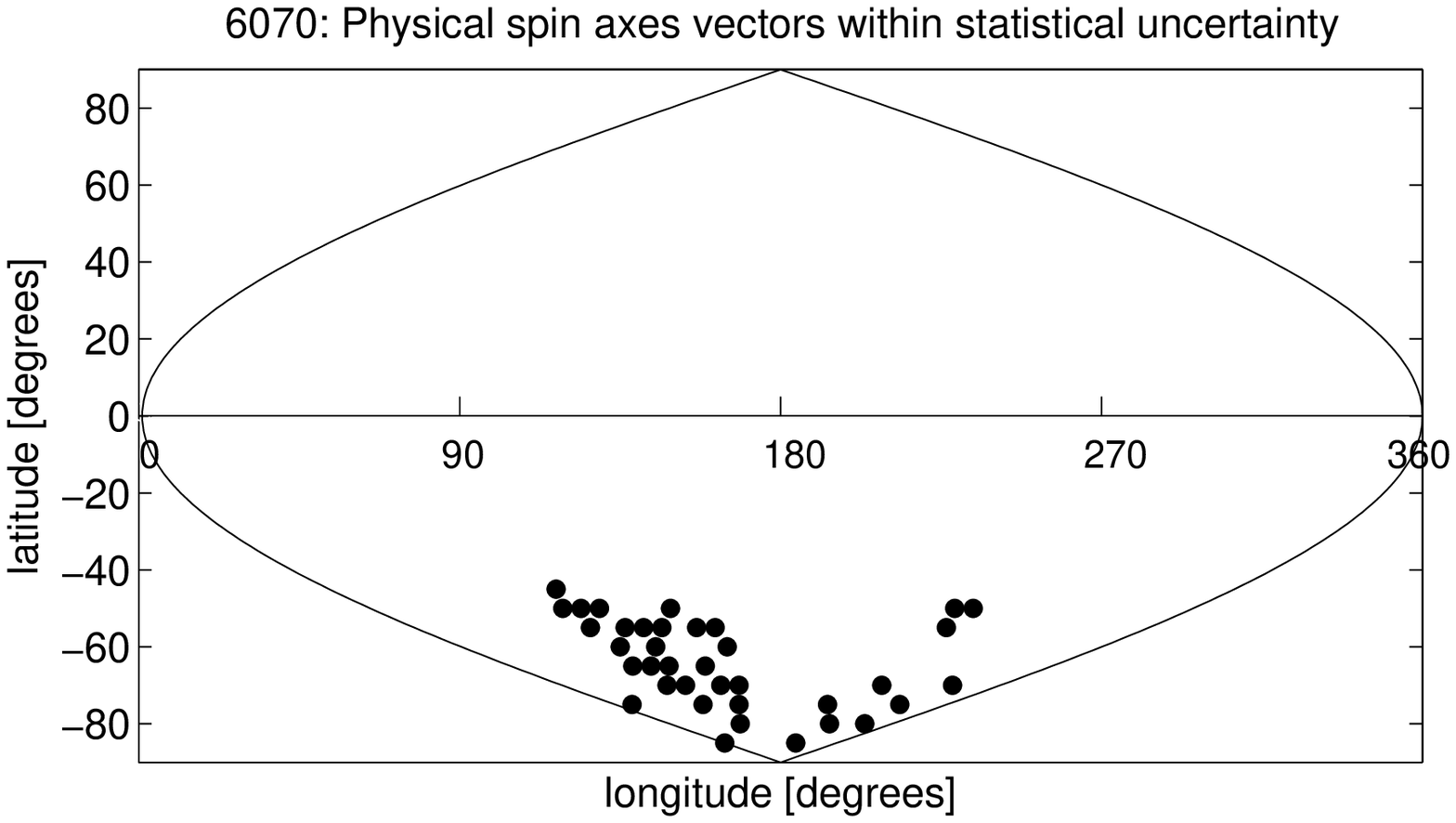}}
\caption{Possible coordinates of spin axes that have feasible physical parameters for (6070) {\it Rheinland}. These constrain the latititude of the spin axis to $-60\pm10$ degrees, where the uncertainty is the standard deviation of the latitude of all feasibile axes.
\label{fig:6070_feasAxis}}
\end{figure}

	Here I use photometric observations from the Wise Observatory taken on 24 nights during four different apparitions between 2009 to 2014 (Fig.~\ref{fig:6070_obs}). Part of this dataset was previously published by Pravec et al. (2010) and Vokrouhlick{\'y} et al. (2011; Table~\ref{tab:ObsCircum1}). In addition to non-published data from the Wise Observatory this photometric dataset was expanded by 30 nights of sparse photometric data collected by the Catalina Sky Survey and the U.S. Naval Observatory. The high brightness of 6070 allowed us to measure small features on its lightcurves with high SNR (Fig.~\ref{fig:6070_LC}). Using the inverse lightcurve metod, and after rejecting unphysical cases as described above, 37 possible spin axis solutions were derived within the uncertainty range that corresponds to $3\sigma$ above the global minimum (Fig.~\ref{fig:6070_axis}$-$~\ref{fig:6070_feasAxis}). The derived spin axes correspond to a sidereal rotation period of $4.273715\pm0.000003$ h, and a retrograde sense of rotation with a latitude of $-60^o\pm10^o$. These values are consistence with the values obtained by Vokrouhlick{\'y} et al. (2011).

\begin{figure}
\centerline{\includegraphics[width=8.5cm]{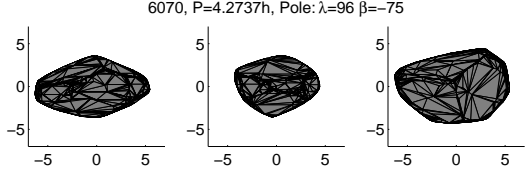}}
\caption{A typical shape model for (6070) {\it Rheinland} from the lightcurve inversion analysis. The three views are from equatorial level (left and center) and pole-on (right). All derived shape models present a pointy side compared to a more circular area on the opposite side. The pointy southern hemisphere appear in 33 out of the 37 possible models.
\label{fig:6070_shape2}}
\end{figure}

\begin{figure}
\centerline{\includegraphics[width=8.5cm]{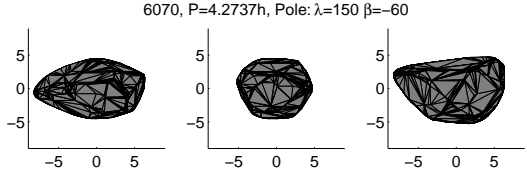}}
\caption{Another possible shape model for (6070) {\it Rheinland}. In this example the shouthren hemisphere does not present a pointy shape. 4 out of 37 models have similar shape to the shape presented here.
\label{fig:6070_shape1}}
\end{figure}

	The shape models constrained by all 37 spin axis solutions are very similar; the shape is highly asymmetric with one pointy side compared to a more rounded area at the opposite side (an example is shown in Fig.~\ref{fig:6070_shape2}) along the long axis of the asteroid ({\it a}). The shortest axis also seems to be asymmetric with a more pointy shouthern hemisphere compared to the northern one. However, in four out of the 37 cases the two hemispheres look symmetric and less pointy (example in Fig.~\ref{fig:6070_shape1}). The southern hemisphere of (6070) {\it Rheinland} will be in opposition again on March 2015, hopefully allowing us to remove this ambiguity. Matching a triaxial ellipsoid to the shape models results with an $a/b$ ratio of $1.38\pm0.06$ and a $b/c$ ratio of $1.2\pm0.2$.

\begin{figure}
\centerline{\includegraphics[width=8.5cm]{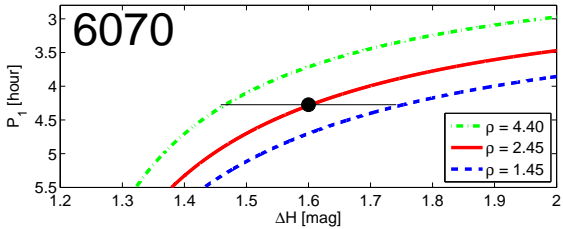}}
\caption{The measured rotation period of (6070) {\it Rheinland} and its size/mass ratio with its secondary (black circle) compared to the values obtained by the model of angular momentum conservation (lines). The three lines represent the mean and two extreme cases with the stated density ($\rho$) values. These include the extreme cases of the triaxial axes of the shape model as derived from our analysis. The error on the Y-axis is negligible. The asymptotic nature of the model makes the uncertainty of the density highly dependent on the error of the $\Delta H$ value for asteroid pairs with low $\Delta H$. Therefore, the derived density range for 6070 is too large to reveal the structure of the asteroid. Decreasing the uncertainty of the absolute magnitude of the primary and secondary is a first step to further constrain the density of 6070.
\label{fig:6070_dens}}
\end{figure}

	Constraining the rotational-fission model by the measured parameters results with a very large range of density values (1.5 to 4.4 $gr ~ cm^{-3}$; Fig.~\ref{fig:6070_dens}). This is because the asymptotic nature of the model makes the uncertainty of the density highly dependent on the error of the $\Delta H$ value for asteroid pairs with low $\Delta H$. Therefore, the derived density range for 6070 is too large to reveal the structure of the asteroid. However, regardless the error range of $\Delta H$, the obtained density value ($2.5 ~ gr ~ cm^{-3}$) is lower than the value of ordinary chondrites ($\sim3.3 ~ gr ~ cm^{-3}$) that 6070 is composed of. As in the previous cases, this density value is consistent with the notion of rubble pile asteroids disrupted by the rotational-fission mechanism.
	
\subsection{(7343) Ockeghem}
\label{sec:ockeghem}
	Asteroid (7343) {\it Ockeghem} was separated from its secondary (154634) {\it 2003 XX38} over 800 kyrs ago (Pravec et al. 2010). Both components present reflectance spectrum of a S-type supporting a common origin (Duddy et al. 2012). In addition, the relatively fast rotation of 7343 (3.75 hours) is correlated with the small size of its secondary (Pravec et al. 2010) which is consistent with the rotational-fission mechanism.

\begin{figure}
\centerline{\includegraphics[width=8.5cm]{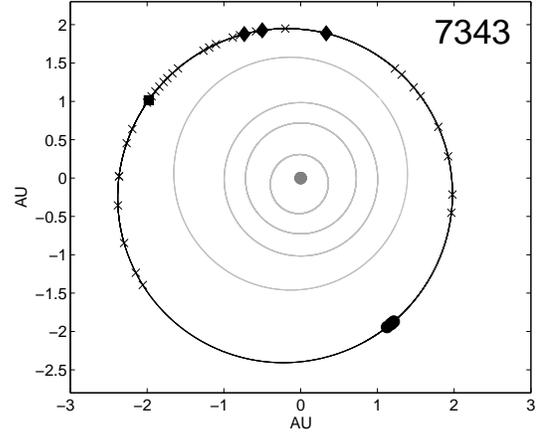}}
\caption{The locations on the orbit of (7343) {\it Ockeghem} where it was observed for this study. 7343 was observed on July 2009 (circles), March 2011 (squares) and November 2013 to January 2014 (diamonds). Data taken by large-scale surveys (x) were collected between January 19, 2005 to June 13, 2011.
\label{fig:7343_obs}}
\end{figure}

\begin{figure}
\centerline{\includegraphics[width=8.5cm]{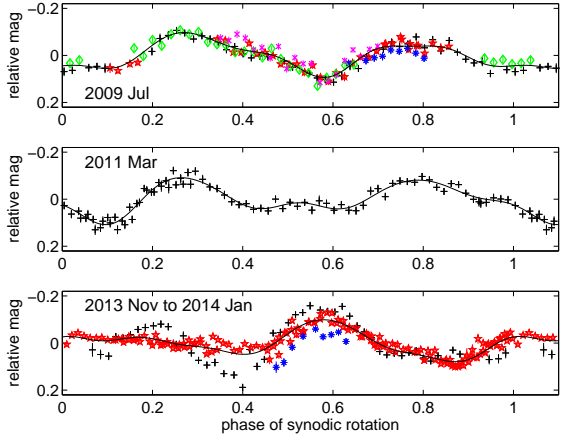}}
\caption{Folded lightcurves of (7343) {\it Ockeghem} from three apparitions. The epoch is $T_0=2455034 JD$. Different markers and colors represent photometric data from different observing nights (see Table~\ref{tab:ObsCircum1}). The synodic rotation period is $3.755\pm0.007$ hours. The lightcurve from November 4, 2013 (bottom panel, black crosses) has the same periodicity as the other lightcurves but it looks different since it was obtained when the asteroid was at the other side of the opposition, meaning the observation took place in a significantly different aspect angle.
\label{fig:7343_LC}}
\end{figure}

\begin{figure}
\centerline{\includegraphics[width=8.5cm]{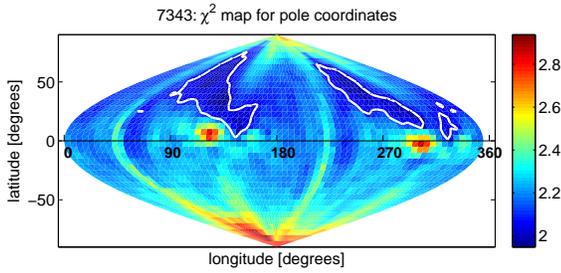}}
\caption{The $\chi^2$ values for spin axis solutions with a period of 3.7553 hours on a longitude-latitude plane for (7343) {\it Ockeghem}. The white line marks the uncertainty range that corresponds to $1\sigma$ above the global minimum. The lightcurve inversion analysis gives different $\chi^2$ planes within the uncertainty of this $\chi^2$ plane when other sidereal periods are used.
\label{fig:7343_axis}}
\end{figure}

\begin{figure}
\centerline{\includegraphics[width=8.5cm]{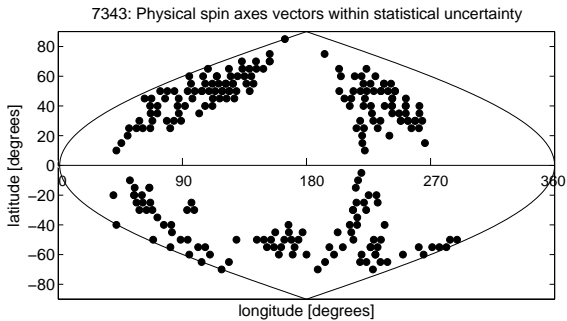}}
\caption{Possible coordinates of spin axes within the uncertainty range that corresponds to $1\sigma$ above the global minimum with feasible physical parameters for (7343) {\it Ockeghem}. The sense of rotation and exact spin axis coordinates could not be derived from a solution that matches the measurements significantly better.
\label{fig:7343_feasAxis}}
\end{figure}

	I observed 7343 at Wise Observatory on 11 nights during 3 apparitions from 2009 to 2014 (Fig.~\ref{fig:7343_obs}). The photometric dataset was increased with measurements from 32 nights collected by the Catalina Sky Survey between 2005 and 2011. While the lightcurve was covered entirely during the rotation period at three different apparitions (Fig.~\ref{fig:7343_LC}), the lightcurve inversion method failed to result with a highly constrained spin rate and spin axis solutions (an example for one solution is displayed in Fig.~\ref{fig:7343_axis}). Therefore, the results presented here have high formal errors, resulting from the standard deviation of all 234 possible solutions (Fig.~\ref{fig:7343_feasAxis}).  $\sim60\%$ of the valid solutions have a prograde rotation, therefore, a favorite sense of rotation cannot be determined as well.

\begin{figure}
\centerline{\includegraphics[width=8.5cm]{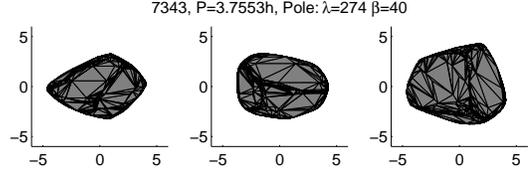}}
\caption{A typical example of a shape model with prograde rotation for (7343) {\it Ockeghem} from the lightcurve inversion analysis. The three views are from equatorial level (left and center) and pole-on (right).
\label{fig:7343_shape1}}
\end{figure}

\begin{figure}
\centerline{\includegraphics[width=8.5cm]{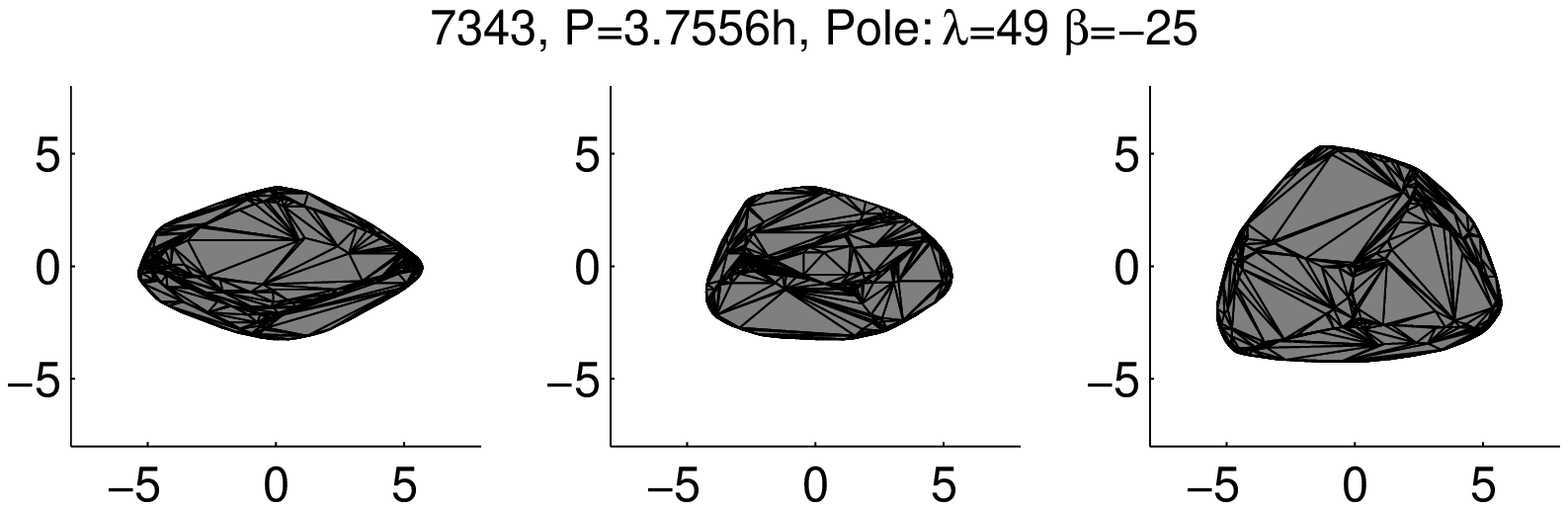}}
\caption{A typical example of a shape model with retrograde rotation for (7343) {\it Ockeghem} from the lightcurve inversion analysis. The three views are from equatorial level (left and center) and pole-on (right).
\label{fig:7343_shape2}}
\end{figure}

	However, the shape models resulted from all of these solutions converge into a non-elongated ($a/b=1.11\pm0.04$, $b/c=1.3\pm0.2$), regular polygon-like shape with pointy corners (e.g. Fig.~\ref{fig:7343_shape1}$-$~\ref{fig:7343_shape2}). In addition, fitting the rotational-fission model to the derived data, the density is constrained to 1.0 to 1.8 $gr ~ cm^{-3}$ which might be considered too low for the density of ordinary chondrite asteroids with a rubble pile structure. Additional measurements in the future will improve the constraints on the rotation and shape of 7343, and will determine if this mismatch is due to flaws in the analysis code, in the understanding of 7343's composition, or in the understanding of the rotational-fission mechanism.

\begin{figure}
\centerline{\includegraphics[width=8.5cm]{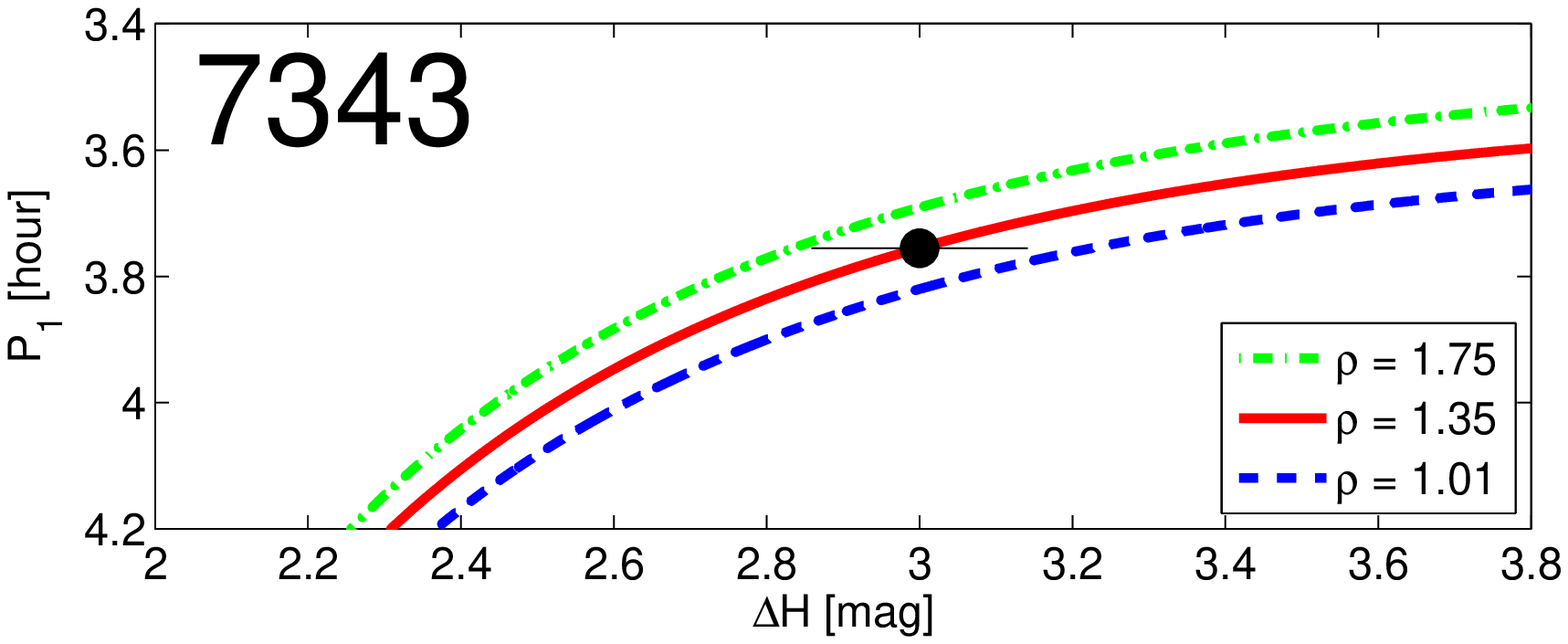}}
\caption{The measured rotation period of (7343) {\it Ockeghem} and its size/mass ratio with its secondary (black circle) compared to the values obtained by the model of angular momentum conservation (lines). The three lines represent the mean and two extreme cases with the stated density ($\rho$) values. These include the extreme cases of the triaxial axes of the shape model as derived from our analysis. The error on the Y-axis is negligible. The derived density range of (1.01 to 1.75 $gr ~ cm^{-3}$) is relatively small for an S-complex object; while this can be interpret as an example of a very high pororsity and shuttered structure, it might be due to a wrong shape model matched by the analysis method, since the photometric coverage of this asteroid is relatively limited.
\label{fig:7343_dens}}
\end{figure}

\section{Summary and discussion}
\label{sec:discussion}

	I collected photometric data on six asteroids in pairs during multiple apparitions and viewing geometries in order to derive the spin axes and shape models of these bodies. I used the lightcurve inversion method to fit the photometric measurements to a set of possible physical solutions and present the average results with the standard deviation as the uncertainty range of the sidereal period, spin axis longitude and latitude, and shape triaxial ratios $a/b$ and $b/c$.

	A simple model, that is based on the assumption of angular momentum conservation (Pravec \& Harris 2007, Pravec et al. 2010; Appendix A), is matched to the measured parameters of spin, shape and mass ratio between the pairÕs members. This match makes it possible to derive the density of the primary member of each of the studied pairs as was never done in previous studies. In three cases (2110, 3749, 5026) the range for possible densities were compatible with the expected value of a rubble pile asteroid. In one case (6070) the range of possible densities was too wide to infer any significant conclusion, even though the expected density of a rubble pile was within the derived range. In one case (7343) the density of the asteroid seemed too low although still possible. This result supports our understanding of the rotational-fission as a mechanism that can modify and split up a rubble pile structured body. This also increases our confidence in the used model and in its basic assumption of angular momentum conservation as a valid description of the rotational-fission mechanism.
    
	Deriving the spin axis latitude of the pairs, helps us determine if they indeed were split by the YORP effect, that can spin-up and align spin axes of asteroids, or by collisions that should result with a random distribution and tumbling rotations. First, the dynamical age of an asteroid pair should be compared to the timescale of the YORP effect in order to assure that the current rotation state was {\bf not} modified by the YORP effect {\bf after} the rotational-fission event took place. The change in the rotation period of the asteroid $\Delta P$ due to the YORP effect since its formation by the rotational-fission event, was calculated by applying Eq.~\ref{eq:yorp} to \ref{eq:dPdt} on the physical values summarized in Table~\ref{tab:PairsParam} to \ref{tab:PairsDensity} and multiplying $dP/dt$ by the dynamical age. Table~\ref{tab:timescales} summarizes the dynamical ages of the asteroids (as appear in Table~\ref{tab:PairsParam}), the change in their spin rate $d \omega /dt$, their YORP timescale $\tau_{yorp}$, and $\Delta P$ since their formation. The derived values show that the YORP timescale is in one to four orders of magnitude longer than the dynamical age of all six asteroids. If the YORP effect currently spins-up these asteroids they only lost a few minutes or less of their rotation periods since their formation ($\Delta P<2~min$ for 5 of the asteroids, and $\Delta P\sim19~min$ for asteroid 44612; Table~\ref{tab:timescales}). Therefore, we can conclude that the current spin state of the observed asteroids were not modified significantly by the YORP effect following the fission event.

\begin{deluxetable*}{cccccc}
\tablecolumns{6}
\tablewidth{0pt}
\tablecaption{Relevant timescales for the observed asteroids}
\tablehead{
\colhead{Asteroid} &
\colhead{Dynamical Age} &
\colhead{$\frac{d \omega}{dt}$} &
\colhead{$\tau_{yorp}$} &
\colhead{$\Delta P$} &
\colhead{$\tau_{damp}$} \\
\colhead{} &
\colhead{[Myrs]} &
\colhead{[$rad~yr^{-2}$]} &
\colhead{[Myrs]} &
\colhead{[minutes]} &
\colhead{[Myrs]} \\
}
\startdata
2110 & $>1.6$ & $6x10^{-05}$ & 290 & -1.1 & $0.2-1.8$ \\
3749 & $0.28^{-0.025}_{+0.045} $ & $5x10^{-05}$ & 400 & -0.12 & $0.1-1.0$ \\
5026 & $0.018\pm0.001$ & $3x10^{-05}$ & 420 & -0.01 & $0.3-2.8$ \\
6070 & $0.017\pm0.0005$ & $6x10^{-05}$ & 210 & -0.02 & $0.5-4.7$ \\
7343 & $>0.8$	 & $14x10^{-05}$ & 100 & -1.76 & $0.7-7.1$ \\
44612 & $>1.6$ & $45x10^{-05}$ & 25 & -18.88 & $4.5-45.1$
\enddata
\tablenotetext{}{Values used in Eq.~\ref{eq:yorp} to \ref{eq:damping}: $F=10^{14}~kg~km~s^{-2}$ (Jacobson et al. 2014). $Y=0.008$ which is the average value for the 7 asteroids $Y$ was measured for (Rozitis \& Green 2013). $\mu Q = 5x10^{12}$ to $5x10^{13}$, $K_3^2=0.03$ (Harris 1994). $a$, $e$ and $R$ are summarized in Table~\ref{tab:PairsParam}. $\rho$ and $\omega=\frac{2\pi}{P}$ are derived by this study (Table~\ref{tab:Results}$-$\ref{tab:PairsDensity}).}
\label{tab:timescales}
\end{deluxetable*}

	Second, the dynamical age of an asteroid pair should be compared to the damping timescale $\tau_{damp}$ of a possible tumbling rotation. $\tau_{damp}$ is calculated by using Eq.~\ref{eq:damping} and values rationalized by Harris (1994; Table~\ref{tab:timescales}). The uncertainty has one order of magnitude range, which originates from the guessed rigidity and quality factor of the asteroids. The $\tau_{damp}$ of three objects have the same order of magnitude as their dynamical ages, therefore, if these objects were formed by a collision, their tumbling motion would have relaxed since their formation. However, for the smallest object, 44612, and for the youngest pairs (5026 and 6070), the dynamical ages are significantly shorter than $\tau_{damp}$. Therefore, we can revoke the idea that these three asteroids were formed by a collision, since they do not present any tumbling rotation. Asteroid 2110 should be added to these three asteroids since it is the pair of 44612, and they both formed by the same mechanism.

	Not only asteroid 44612 does not present a tumbling rotation, this asteroid and its partner 2110 share the same sense of rotation and the direction of their spin axes are within the uncertainty range. This result should be considered as an additional evidence that the YORP effect caused the rotational-fission since the disruption is orthogonal to the spin axis, therefore the fission should not modify the direction of the spin axis.

\begin{figure}
\centerline{\includegraphics[width=8.5cm]{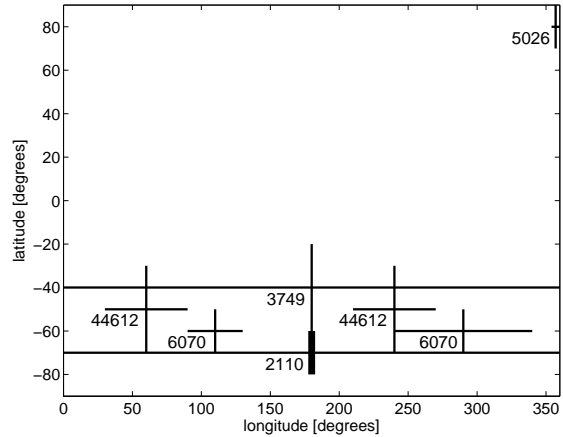}}
\caption{The distribution of longitude and latitude of the spin axis vectors of five of the observed asteroid pairs (values of 7343 were omitted since their uncertainty include almost the entire graph). Three asteroids (2110, 5026, 6070) resulted with polar-directed spin axes and not equatorial-directed spin axes and the three others (3749, 7343, 44612) with ambiguous results. This supports the notion that the YORP effect is the mechanism that spun-up the asteroids what eventually resulted in their fission to asteroid pairs.
\label{fig:allpoles}}
\end{figure}

	Checking the entire sample, for three of the studied cases (2110, 5026, 6070), the spin axis is indeed pointed towards the general direction of the north (5026) or south (2110, 6070) pole of the asteroid (Fig.~\ref{fig:allpoles}). This result gives a supporting argument to the notion that the fission of these pairs occurred after the YORP effect modified and aligned the spin axes of these asteroids. Since the thermal reemission from the asteroid surface gives momentum to the asteroid in a normal and parallel vectors compared to the spin axis (Vokrouhlick{\'y} et al. 2003), an evidence for the spin axis modification by the YORP effect serves as an evidence that the spin rate of the asteroid was modified by the YORP effect as well, what eventually caused the disruption of the asteroid. The other three cases (3749, 7343, 44612) resulted in ambiguous values that include polar and equatorial directed spin axes, so conclusions cannot be drawn from these three asteroids.
	
	The study presented here begins the long task of assembling the necessary data to better understand the rotational fission process. Observing more asteroid pairs and collecting their photometric data in multiple observing geometries will give a statistical point of view on the spin and shapes of asteroid pairs. This will improve our understanding of the rotational-fission mechanism and about evolution of asteroids in general. Using the model described in the paper, the pairs give us a unique opportunity to increase the number of known density values and the opportunity to constrain these values with the known asteroidal structure of ``rubble pile". Furthermore, reducing the uncertainties of the derived solutions could even produce a shape model with enough detail to reveal the place on the surface where the fission took place.

\appendix
\label{sec:ApendixA}

	Pravec and Harris (2007) and Pravec et al. (2010) presented a rotational-fission model based on the conservation of angular momentum. The source of angular momentum for two touching bodies is the spin of each object and their orbit around each other. This can be formulated as:

\begin{eqnarray}
\frac{1}{2}I_1\omega_{1ini}^2 + \frac{1}{2}I_2\omega_{2ini}^2 -G\frac{M_1M_2}{2A_{ini}} & = & 
 \frac{1}{2}I_1\omega_{1fin}^2 + \frac{1}{2}I_2\omega_{2fin}^2 -G\frac{M_1M_2}{2A_{fin}}
\label{eq:MomConserv}
\end{eqnarray}

where $I_i$, $\omega_i$, $M_i$ are the moment of inertia around the principal axis, the angular velocity and the mass of the i-th body (primary is 1, secondary is 2), respectively, A is the semi-major axis of the system and G is the gravitational constant. The subscripts ``ini" and ``fin" stands for the initial and final state values of the relevant parameters. To simplify the model we can make the following assumptions:

1. The density and albedo of the two components are equal.

2. The spin of the secondary is constant ($\omega_{2ini} = \omega_{2fin}$). Any change within this value is negligible due to the small size of the secondary.

3. The spin vectors of the components and their mutual orbit are coplanar (according to this study, this assumption is true for 2110-44612).

Because the final state of the two components is total separation, $1/A_{fin}$ is defined as zero, meaning a barely escaping satellite with a parabolic orbit. Therefore, Eq.~\ref{eq:MomConserv} can be expressed as:

\begin{equation}
\frac{1}{2}I_1\omega_{1ini}^2 -G\frac{M_1M_2}{2A_{ini}} =\frac{1}{2}I_1\omega_{1fin}^2 
\label{eq:MomConservShort}
\end{equation}

Some parameters can be replaced by known parameters or observed ratios. These are the moment of inertia of the primary $I_1$, the mass of the primary $M_1$ and the secondary $M_2$:

\begin{equation}
I_1 = \frac{M_1}{5}(a_1^2+b_1^2)
\label{eq:MomConservShort}
\end{equation}

\begin{equation}
M_1 = \frac{4\pi}{3}\rho(a_1b_1c_1)
\label{eq:MomConservShort}
\end{equation}

\begin{equation}
M_2 = qM_1
\label{eq:MomConservShort}
\end{equation}

where $a_1$, $b_1$, $c_1$ are the physical axes of an equal mass ellipsoid ($a_1 \geq b_1 \geq c_1$), $\rho$ is the density and $q$ is the mass ratio between the components. After the substitutions, we get:

\begin{equation}
\omega_{1fin}^2 = \omega_{1ini}^2 - \frac{20\pi\rho G\frac{a_1}{b_1}\frac{c_1}{b_1}}{3[1+(\frac{a_1}{b_1})^2]\frac{A_{ini}}{b_1}}q
\label{eq:MomConservShort}
\end{equation}

The triaxial ratio of the pair primaries ($a_1$/$b_1$ and $b_1$/$c_1$) is calculated from the derived shape models. The initial semi-major axis $A_{ini}$, normalized by the intermediate axis of the primary $b_1$, is of order unity for a contact binary.

The initial spin of the primary $\omega_{1ini}$ is calculated from the total angular momentum, $L_1+L_2+L_{orb}$, which is the primary and secondary rotational angular momentum and the orbital angular momentum, respectively. This value is normalized by the angular momentum of an equivalent sphere $L_{eqsph}$, spinning at the critical spin rate:

\begin{equation}
\alpha_L = \frac{L_1+L_2+L_{orb}}{L_{eqsph}}
\label{eq:MomConservShort}
\end{equation}

The parameters in the above formula can be replaced by:

\begin{equation}
L_1 = \frac{M}{5(1+q)}(a_1^2+b_1^2)\omega_1
\label{eq:MomConservShort}
\end{equation}

\begin{equation}
L_2 = \frac{qM}{5(1+q)}(a_2^2+b_2^2)\omega_2
\label{eq:MomConservShort}
\end{equation}

\begin{equation}
L_{orb} = \frac{qM}{(1+q)^2}\sqrt{GMA_{ini}(1-e^2)}
\label{eq:MomConservShort}
\end{equation}

\begin{equation}
L_{eqsph} = \frac{2}{5}M(a_1b_1c_1)^{2/3}(1+q)^{2/3}\omega_{csph}
\label{eq:MomConservShort}
\end{equation}

when $M = M_1 + M_2$ and the critical spin rate of a sphere, $\omega_{csph}$, is determined by its density as:

\begin{equation}
\omega_{csph} = \sqrt{\frac{4\pi}{3}G\rho}
\label{eq:MomConservShort}
\end{equation}

Eventually, the initial spin of the primary $\omega_{1ini}$ can be formulated as follows:

\begin{eqnarray}
\omega_{1ini} = \frac{1}{(\frac{a_1^2}{b_1^2}+1)}[2\alpha_L(\frac{a_1}{b_1}\frac{c_1}{b_1})^{2/3}(1+q)^{5/3}\sqrt{\frac{4\pi}{3}G\rho} & 
-\frac{5q}{q+1}\sqrt{\frac{4\pi}{3}G\rho\frac{a_1}{b_1}\frac{c_1}{b_1}(1+q)\frac{A_{ini}}{b_1}(1-e^2)} 
-q(\frac{a_2^2}{b_2^2}+1)\omega_2]
\label{eq:MomConservShort}
\end{eqnarray}

\acknowledgments

I thank the reviewers for their useful comments and suggestions that improve the paper's arguments. I am grateful to Josef {\v D}urech and Josef Hanu{\v s} for providing their lightcurve inversion code and guiding me through it. I would like to acknowledge the AXA research fund for their generous postdoctoral fellowship. Richard Binzel, Avishay Gal-Yam, Noah Brosch and Dina Prialnik were all generous for giving the needed support during the seven long years of observations. I am deeply thankful to the Wise Observatory staff for their continuous help and generous time allocation. And finally, Talia Jacobi is praised for her reconciliation with the toll of astronomical observations.


\begin{thebibliography}{}

\bibitem[Asphaug and Scheeres(1999)]{1999Icar..139..383A} Asphaug, E., Scheeres, D.~J.\ 1999.\ NOTE: Deconstructing Castalia: Evaluating a Postimpact State.\ Icarus 139, 383-386.

\bibitem[Binzel et al.(1996)]{1996Sci...273..946B} Binzel, R.~P., Bus, S.~J., Burbine, T.~H., Sunshine, J.~M.\ 1996.\ Spectral Properties of Near-Earth Asteroids: Evidence for Sources of Ordinary Chondrite Meteorites.\ Science 273, 946-948.

\bibitem[Binzel et al.(2010)]{2010Natur.463..331B} Binzel, R.~P., Morbidelli, A., Merouane, S., DeMeo, F.~E., Birlan, M., Vernazza, P., Thomas, C.~A., Rivkin, A.~S., Bus, S.~J., Tokunaga, A.~T.\ 2010.\ Earth encounters as the origin of fresh surfaces on near-Earth asteroids.\ Nature 463, 331-334.

\bibitem[Bottke et al.(2006)]{2006AREPS..34..157B} Bottke, W.~F., Jr.,  Vokrouhlick{\'y}, D., Rubincam, D.~P., Nesvorn{\'y}, D.\ 2006.\ The  Yarkovsky and Yorp Effects: Implications for Asteroid Dynamics.\ Annual  Review of Earth and Planetary Sciences 34, 157-191.

\bibitem[Bowell et al.(1989)]{1989aste.conf..524B} Bowell, E., Hapke, B.,  Domingue, D., Lumme, K., Peltoniemi, J., Harris, A.~W.\ 1989.\ Application  of photometric models to asteroids.\ Asteroids II 524-556.

\bibitem[Brosch et  al.(2008)]{2008Ap&SS.314..163B} Brosch, N., Polishook, D., Shporer, A., Kaspi, S., Berwald, A., Manulis, I.\ 2008.\ The Centurion 18 telescope of the Wise Observatory.\ Astrophysics and Space Science 314, 163-176.

\bibitem[Brunetto et al.(2006)]{2006Icar..184..327B} Brunetto, R., Vernazza, P., Marchi, S., Birlan, M., Fulchignoni, M., Orofino, V., Strazzulla, G.\ 2006.\ Modeling asteroid surfaces from observations and irradiation experiments: The case of 832 Karin.\ Icarus 184, 327-337. 

\bibitem[Burns and Safronov(1973)]{1973MNRAS.165..403B} Burns, J.~A., Safronov, V.~S.\ 1973.\ Asteroid nutation angles.\ Monthly Notices of the Royal Astronomical Society 165, 403.

\bibitem[Carry(2012)]{2012P&SS...73...98C} Carry, B.\ 2012.\ Density of asteroids.\ Planetary and Space Science 73, 98-118.

\bibitem[Clark et al.(2002)]{2002aste.conf..585C} Clark, B.~E., Hapke, B., Pieters, C., Britt, D.\ 2002.\ Asteroid Space Weathering and Regolith Evolution.\ Asteroids III 585-599.

\bibitem[DeMeo et al.(2009)]{2009Icar..202..160D} DeMeo, F.~E., Binzel, R.~P., Slivan, S.~M., Bus, S.~J.\ 2009.\ An extension of the Bus asteroid taxonomy into the near-infrared.\ Icarus 202, 160-180.

\bibitem[Duddy et al.(2012)]{2012A&A...539A..36D} Duddy, S.~R., Lowry, S.~C., Wolters, S.~D., Christou, A., Weissman, P., Green, S.~F., Rozitis, B.\ 2012.\ Physical and dynamical characterisation of the unbound asteroid pair 7343-154634.\ Astronomy and Astrophysics 539, A36.

\bibitem[Duddy et al.(2013)]{2013MNRAS.429...63D} Duddy, S.~R., Lowry, S.~C., Christou, A., Wolters, S.~D., Rozitis, B., Green, S.~F., Weissman, P.~R.\ 2013.\ Spectroscopic observations of unbound asteroid pairs using the WHT.\ Monthly Notices of the Royal Astronomical Society 429, 63-74.

\bibitem[Durda et al.(2004)]{2004Icar..170..243D} Durda, D.~D., Bottke, W.~F., Enke, B.~L., Merline, W.~J., Asphaug, E., Richardson, D.~C., Leinhardt, Z.~M.\ 2004.\ The formation of asteroid satellites in large impacts: results from numerical simulations.\ Icarus 170, 243-257.

\bibitem[Durech et al.(2008)]{2008A&A...489L..25D} {\v D}urech, J., and 11 colleagues 2008.\ Detection of the YORP effect in asteroid (1620) Geographos.\ Astronomy and Astrophysics 489, L25-L28.

\bibitem[Durech et  al.(2010)]{2010A&A...513A..46D} {\v D}urech, J., Sidorin, V., Kaasalainen, M.\ 2010.\ DAMIT: a database of asteroid models.\ Astronomy and Astrophysics 513, A46.

\bibitem[{\v D}urech et al.(2012)]{2012A&A...547A..10D} {\v D}urech, J., and 30 colleagues 2012.\ Analysis of the rotation period of asteroids (1865) Cerberus, (2100) Ra-Shalom, and (3103) Eger - search for the YORP effect.\ Astronomy and Astrophysics 547, A10.

\bibitem[Fujiwara et al.(2006)]{2006Sci...312.1330F} Fujiwara, A., and 21 colleagues 2006.\ The Rubble-Pile Asteroid Itokawa as Observed by Hayabusa.\ Science 312, 1330-1334. 

\bibitem[Giblin et al.(1998)]{1998Icar..134...77G} Giblin, I., Martelli, G., Farinella, P., Paolicchi, P., Di Martino, M., Smith, P.~N.\ 1998.\ The Properties of Fragments from Catastrophic Disruption Events.\ Icarus 134, 77-112.

\bibitem[Hanu{\v s} et  al.(2011)]{2011A&A...530A.134H} Hanu{\v s}, J., and 14 colleagues 2011.\ A study of asteroid pole latitude distribution based on an extended set of shape models derived by the lightcurve inversion method.\ Astronomy and Astrophysics 530, A134.

\bibitem[Hanu{\v s} et al.(2013)]{2013A&A...559A.134H} Hanu{\v s}, J., and 13 colleagues 2013.\ An anisotropic distribution of spin vectors in asteroid families.\ Astronomy and Astrophysics 559, A134.

\bibitem[Harris(1994)]{1994Icar..107..209H} Harris, A.~W.\ 1994.\ Tumbling asteroids.\ Icarus 107, 209.

\bibitem[Jacobson and Scheeres(2011)]{2011Icar..214..161J} Jacobson, S.~A.,  Scheeres, D.~J.\ 2011.\ Dynamics of rotationally fissioned asteroids:  Source of observed small asteroid systems.\ Icarus 214, 161-178.

\bibitem[Jacobson et al.(2014)]{2014MNRAS.439L..95J} Jacobson, S.~A., Marzari, F., Rossi, A., Scheeres, D.~J., Davis, D.~R.\ 2014.\ Effect of rotational disruption on the size-frequency distribution of the Main Belt asteroid population.\ Monthly Notices of the Royal Astronomical Society 439, L95-L99.

\bibitem[Jewitt et al.(2010)]{2010Natur.467..817J} Jewitt, D., Weaver, H., Agarwal, J., Mutchler, M., Drahus, M.\ 2010.\ A recent disruption of the main-belt asteroid P/2010A2.\ Nature 467, 817-819. 

\bibitem[Jewitt et al.(2013)]{2013ApJ...778L..21J} Jewitt, D., Agarwal, J., Weaver, H., Mutchler, M., Larson, S.\ 2013.\ The Extraordinary Multi-tailed Main-belt Comet P/2013 P5.\ The Astrophysical Journal 778, L21. 

\bibitem[Jewitt et al.(2014)]{2014ApJ...784L...8J} Jewitt, D., Agarwal, J., Li, J., Weaver, H., Mutchler, M., Larson, S.\ 2014.\ Disintegrating Asteroid P/2013 R3.\ The Astrophysical Journal 784, L8. 

\bibitem[Kaasalainen and Torppa(2001)]{2001Icar..153...24K} Kaasalainen,  M., Torppa, J.\ 2001.\ Optimization Methods for Asteroid Lightcurve  Inversion. I. Shape Determination.\ Icarus 153, 24-36.

\bibitem[Kaasalainen et al.(2001)]{2001Icar..153...37K} Kaasalainen, M.,  Torppa, J., Muinonen, K.\ 2001.\ Optimization Methods for Asteroid  Lightcurve Inversion. II. The Complete Inverse Problem.\ Icarus 153, 37-51.

\bibitem[Kaasalainen et al.(2007)]{2007Natur.446..420K} Kaasalainen, M., {\v D}urech, J., Warner, B.~D., Krugly, Y.~N., Gaftonyuk, N.~M.\ 2007.\ Acceleration of the rotation of asteroid 1862 Apollo by radiation torques.\ Nature 446, 420-422.

\bibitem[Landolt(1992)]{1992AJ....104..340L} Landolt, A.~U.\ 1992.\ UBVRI photometric standard stars in the magnitude range 11.5-16.0 around the celestial equator.\ The Astronomical Journal 104, 340-371.

\bibitem[Lowry et al.(2007)]{2007Sci...316..272L} Lowry, S.~C., and 10 colleagues 2007.\ Direct Detection of the Asteroidal YORP Effect.\ Science 316, 272.

\bibitem[Mainzer et al.(2011)]{2011ApJ...741...90M} Mainzer, A., and 12 colleagues 2011.\ NEOWISE Studies of Spectrophotometrically Classified Asteroids: Preliminary Results.\ The Astrophysical Journal 741, 90.

\bibitem[Marchis et al.(2008)]{2008IAUC.8928....4M} Marchis, F., and 13 colleagues 2008a.\ (3749) Balam.\ International Astronomical Union Circular 8928, 4.

\bibitem[Marchis et al.(2008)]{2008Icar..195..295M} Marchis, F., Descamps, P., Berthier, J., Hestroffer, D., Vachier, F., Baek, M., Harris, A.~W., Nesvorn{\'y}, D.\ 2008b.\ Main belt binary asteroidal systems with eccentric mutual orbits.\ Icarus 195, 295-316. 

\bibitem[Margot et al.(2002)]{2002Sci...296.1445M} Margot, J.~L., Nolan, M.~C., Benner, L.~A.~M., Ostro, S.~J., Jurgens, R.~F., Giorgini, J.~D., Slade, M.~A., Campbell, D.~B.\ 2002.\ Binary Asteroids in the Near-Earth Object Population.\ Science 296, 1445-1448.

\bibitem[Marzari et al.(2011)]{2011Icar..214..622M} Marzari, F., Rossi, A., Scheeres, D.~J.\ 2011.\ Combined effect of YORP and collisions on the rotation rate of small Main Belt asteroids.\ Icarus 214, 622-631.

\bibitem[Merline et al.(2002)]{2002IAUC.7827....2M} Merline, W.~J., Close, L.~M., Siegler, N., Dumas, C., Chapman, C., Rigaut, F., Menard, F., Owen, W.~M., Slater, D.~C.\ 2002.\ S/2002 (3749) 1.\ International Astronomical Union Circular 7827, 2.

\bibitem[Moskovitz(2012)]{2012Icar..221...63M} Moskovitz, N.~A.\ 2012.\ Colors of dynamically associated asteroid pairs.\ Icarus 221, 63-71.

\bibitem[Paolicchi et al.(2002)]{2002aste.conf..517P} Paolicchi, P., Burns, J.~A., Weidenschilling, S.~J.\ 2002.\ Side Effects of Collisions: Spin Rate Changes, Tumbling Rotation States, and Binary Asteroids.\ Asteroids III 517-526. 

\bibitem[Polishook and Brosch(2008)]{2008Icar..194..111P} Polishook, D.,  Brosch, N.\ 2008.\ Photometry of Aten asteroids, More than a handful of binaries.\ Icarus 194, 111-124.

\bibitem[Polishook and Brosch(2009)]{2009Icar..199..319P} Polishook, D.,  Brosch, N.\ 2009.\ Photometry and spin rate distribution of small-sized  main belt asteroids.\ Icarus 199, 319-332.

\bibitem[Polishook(2011)]{2011MPBu...38...94P} Polishook, D.\ 2011.\ Rotation Period of the ''Asteroid Pair'' (25884) 2000 SQ4.\ Minor Planet Bulletin 38, 94-95.

\bibitem[Polishook et al.(2011)]{2011Icar..212..167P} Polishook, D., Brosch, N., Prialnik, D.\ 2011.\ Rotation periods of binary asteroids with large separations - Confronting the Escaping Ejecta Binaries model with observations.\ Icarus 212, 167-174.

\bibitem[Polishook(2014)]{2014MPBu...41...49P} Polishook, D.\ 2014.\ Spins, Lightcurves, and Binarity of Eight Asteroid Pairs: 4905, 7745, 8306, 16815, 17288, 26416, 42946, and 74096.\ Minor Planet Bulletin 41, 49-53.

\bibitem[Polishook et al.(2014)]{2014Icar..233....9P} Polishook, D., Moskovitz, N., Binzel, R.~P., DeMeo, F.~E., Vokrouhlick{\'y}, D., {\v Z}i{\v z}ka, J., Oszkiewicz, D.\ 2014.\ Observations of ``fresh" and weathered surfaces on asteroid pairs and their implications on the rotational-fission mechanism.\ Icarus 233, 9-26.

\bibitem[Pravec and Harris(2007)]{2007Icar..190..250P} Pravec, P., Harris, A.~W.\ 2007.\ Binary asteroid population. 1. Angular momentum content.\ Icarus 190, 250-259. 

\bibitem[Pravec and Vokrouhlick{\'y}(2009)]{2009Icar..204..580P} Pravec,  P., Vokrouhlick{\'y}, D.\ 2009.\ Significance analysis of asteroid pairs.\  Icarus 204, 580-588.

\bibitem[Pravec et al.(2005)]{2005Icar..173..108P} Pravec, P., and 19  colleagues 2005.\ Tumbling asteroids.\ Icarus 173, 108-131.

\bibitem[Pravec et al.(2008)]{2008Icar..197..497P} Pravec, P., and 30 colleagues 2008.\ Spin rate distribution of small asteroids.\ Icarus 197, 497-504.

\bibitem[Pravec et al.(2010)]{2010Natur.466.1085P} Pravec, P., and 25  colleagues 2010.\ Formation of asteroid pairs by rotational fission.\  Nature 466, 1085-1088.

\bibitem[Press et al.(2007)] {} Press, W. H., Teukolsky, S. A., Vetterling, W. T. \& Flannery, B. P. 2007, Numerical Recipes (Cambridge: Cambridge Univ. Press)

\bibitem[Rozitis and Green(2013)]{2013MNRAS.430.1376R} Rozitis, B., Green, S.~F.\ 2013.\ The strength and detectability of the YORP effect in near-Earth asteroids: a statistical approach.\ Monthly Notices of the Royal Astronomical Society 430, 1376-1389.

\bibitem[Rubincam(2000)]{2000Icar..148....2R} Rubincam, D.~P.\ 2000.\  Radiative Spin-up and Spin-down of Small Asteroids.\ Icarus 148, 2-11.

\bibitem[Scheeres(2007)]{2007Icar..189..370S} Scheeres, D.~J.\ 2007.\  Rotational fission of contact binary asteroids.\ Icarus 189, 370-385.
 
\bibitem[Slivan(2002)]{2002Natur.419...49S} Slivan, S.~M.\ 2002.\ Spin vector alignment of Koronis family asteroids.\ Nature 419, 49-51. 

\bibitem[Slivan et al.(2008)]{2008Icar..195..226S} Slivan, S.~M., and 10 colleagues 2008.\ Rotation rates in the Koronis family, complete to H{\ap}11.2.\ Icarus 195, 226-276.

\bibitem[Statler et al.(2013)]{2013Icar..225..141S} Statler, T.~S., Cotto-Figueroa, D., Riethmiller, D.~A., Sweeney, K.~M.\ 2013.\ Size matters: The rotation rates of small near-Earth asteroids.\ Icarus 225, 141-155.

\bibitem[Taylor et al.(2007)]{2007Sci...316..274T} Taylor, P.~A., and 11 colleagues 2007.\ Spin Rate of Asteroid (54509) 2000 PH5 Increasing Due to the YORP Effect.\ Science 316, 274.

\bibitem[Vokrouhlick{\'y}(2009)]{2009ApJ...706L..37V} Vokrouhlick{\'y}, D.\  2009.\ (3749) Balam: A Very Young Multiple Asteroid System.\ The  Astrophysical Journal 706, L37-L40.

\bibitem[Vokrouhlick{\'y} and Nesvorn{\'y}(2008)]{2008AJ....136..280V}  Vokrouhlick{\'y}, D., Nesvorn{\'y}, D.\ 2008.\ Pairs of Asteroids Probably  of a Common Origin.\ The Astronomical Journal 136, 280-290.

\bibitem[Vokrouhlick{\'y} and Nesvorn{\'y}(2009)]{2009AJ....137..111V} Vokrouhlick{\'y}, D., Nesvorn{\'y}, D.\ 2009.\ The Common Roots of Asteroids (6070) Rheinland and (54827) 2001 NQ8.\ The Astronomical Journal 137, 111-117.

\bibitem[Vokrouhlick{\'y} et al.(2003)]{2003Natur.425..147V} Vokrouhlick{\'y}, D., Nesvorn{\'y}, D., Bottke, W.~F.\ 2003.\ The vector alignments of asteroid spins by thermal torques.\ Nature 425, 147-151.

\bibitem[Vokrouhlick{\'y} et al.(2011)]{2011AJ....142..159V}  Vokrouhlick{\'y}, D., and 14 colleagues 2011.\ Spin Vector and Shape of  (6070) Rheinland and Their Implications.\ The Astronomical Journal 142,  159.

\bibitem[Walsh et al.(2008)]{2008Natur.454..188W} Walsh, K.~J., Richardson,  D.~C., Michel, P.\ 2008.\ Rotational breakup as the origin of small binary  asteroids.\ Nature 454, 188-191.

\bibitem[Wolters et al.(2014)]{2014MNRAS.439.3085W} Wolters, S.~D., Weissman, P.~R., Christou, A., Duddy, S.~R., Lowry, S.~C.\ 2014.\ Spectral similarity of unbound asteroid pairs.\ Monthly Notices of the Royal Astronomical Society 439, 3085-3093.




\end{thebibliography}
\end{document}